\documentclass[12pt]{elsarticle}
\usepackage{xcolor}
\usepackage{lineno,hyperref}
\usepackage{amsmath}
\usepackage{dsfont}
\usepackage{subfig}
\modulolinenumbers[5]
\usepackage[a4paper]{geometry}
\makeatletter
\def\ps@pprintTitle{%
 \let\@oddhead\@empty
 \let\@evenhead\@empty
 \def\@oddfoot{}%
 \let\@evenfoot\@oddfoot}
\makeatother

\begin{document}

\begin{frontmatter}

\title{A Primitive Variable Discrete Exterior Calculus Discretization of
Incompressible Navier-Stokes Equations over Surface Simplicial Meshes}


\author[mymainaddress]{Pankaj Jagad}
\author[mymainaddress]{Abdullah Abukhwejah}
\author[mysecondaryaddress]{Mamdouh Mohamed}
\author[mymainaddress]{Ravi Samtaney\corref{mycorrespondingauthor}}
\cortext[mycorrespondingauthor]{Corresponding author}
\ead{ravi.samtaney@kaust.edu.sa}

\address[mymainaddress]{Mechanical Engineering, Physical Science and Engineering Division, King Abdullah University of Science and Technology, Thuwal, Saudi Arabia}
\address[mysecondaryaddress]{Department of Mechanical Design and Production, Faculty of Engineering, Cairo University, Giza, Egypt}

\begin{abstract}
A conservative primitive variable discrete exterior calculus (DEC)
discretization of the Navier-Stokes equations is performed. An existing
DEC method (Mohamed, M. S., Hirani, A. N., Samtaney, R. (2016). Discrete exterior calculus discretization of incompressible Navier-Stokes equations over surface simplicial meshes. Journal of Computational Physics, 312, 175-191) is modified to this end, and
is extended to include the energy-preserving time integration and
the Coriolis force to enhance its applicability to investigate the
late time behavior of flows on rotating surfaces, i.e., that of the
planetary flows. The simulation experiments show second order accuracy
of the scheme for the structured-triangular meshes, and first order
accuracy for the otherwise unstructured meshes. The method exhibits
second order kinetic energy relative error convergence rate with mesh
size for inviscid flows. The test case of flow on a rotating sphere
demonstrates that the method preserves the stationary state, and conserves
the inviscid invariants over an extended
period of time.
\end{abstract}

\begin{keyword}
%\texttt{elsarticle.cls}\sep \LaTeX\sep Elsevier \sep template
%\MSC[2010] 00-01\sep  99-00
DEC\sep primitive variable formulation
\end{keyword}

\end{frontmatter}

%\linenumbers

\section{Introduction\label{sec:Introduction}}

Recently, structure preserving numerical methods that satisfy the mathematical and physical properties
of the governing equations are increasingly becoming popular \cite{marsden2001discrete,hairer2006geometric,arnold2007compatible}. 
One of such methods is discrete exterior calculus (DEC). It uses integral quantities / discrete
forms as the primary unknowns and transforms the continuous partial
differential equation (PDE) system into a discrete algebraic system
such that all errors/approximations occur only at the algebraic level
and not in the approximation of the differential operators of the
PDE \cite{perot2007discrete}. Thus, DEC is known to preserve some of the underlying physical and mathematical
properties of the continuous PDE system. 

DEC is the discrete counterpart of smooth exterior calculus, which
is more natural tool than tensor calculus in numerous situations.
The theory of exterior calculus and differential forms dates back
to the pioneering work of Poincar\'e \cite{poincare1887residus} (who
presented the notion of line and surface integrals), Cartan \cite{cartan1899certaines}
( who formalized the notion of differential forms), and Goursat
\cite{goursat1915certains} ( who noted that the three theorems of
vector calculus, viz. Gauss, Green, and Stokes are all special cases
of a generalized Stokes theorem for differential forms). Subsequently,
there were further theoretical developments (for example, see \cite{abraham1932classical,cartan1946leccons,de1955varietes}).
However, there were limited applications of this theory to solve 
physical problems, until the theory of discrete exterior calculus
was developed (see \cite{hirani2003discrete,desbrun2005discrete,desbrun2008discrete}
and the references therein). As this theory advanced, DEC made its
way to the applications such as computer graphics and computer vision
\cite{desbrun2003discrete,grinspun2006discrete,elcott2007stable,desbrun2008discrete,mullen2009energy,crane2013digital,de2016subdivision}
to mitigate the effects of numerical viscosity that causes detrimental
visual consequences, and fluid flows \cite{hirani2008numerical,mohamed2016discrete,nitschke2017discrete,mohamed2018numerical}
to preserve the physical properties of the governing equations, i.e,
to conserve secondary quantities, such as kinetic energy and/or discretely preserving Kelvin's circulation theorem. Developing
such physically conservative discretizations, for Navier-Stokes
equations for example, is favorable for many physical applications
such as turbulent flows \cite{zhang2002accuracy} and shallow-water simulations \cite{frank2003conservation} to avoid undesirable
numerical artifacts.  

Apart from DEC, some staggered mesh methods have also demonstrated
conservation of secondary (kinetic energy and vorticity) physical
quantities in addition to the primary ones (mass and momentum) \cite{perot2011discrete}.
Nicolaides \cite{nicolaides1989flow} and Hall {\it et al.} \cite{hall1991dual}
developed staggered mesh methods, also known as covolume methods, for
unstructured meshes. The covolume discretization commences by taking
dot product of the momentum equation with the unit normal to each
face of the triangular / tetrahedral elements, reducing the velocity
vector to a scalar flux (equal to the mass flux across the face for
an incompressible flow with constant density) defined on each face; and 
 the pressure is defined at the circumcenter of each
triangular / tetrahedral element. Nicolaides \cite{nicolaides1992direct}
estimated the accuracy of the covolume method to be second order for
a mesh with all its triangular elements having the same circumradii
(i.e., structured-triangular mesh), and to be first order otherwise.
Perot \cite{perot2000conservation} demonstrated the conservation
properties of the covolume method. The divergence form of the Navier-Stokes
equations conserved the momentum and kinetic energy both locally and
globally, and the rotational form conserved the circulation and kinetic
energy locally and globally for both 2D \cite{perot2000conservation}
and 3D \cite{zhang2002accuracy} discretizations. A main advantage
of DEC discretizations is their applicability to simulate flows over
curved surfaces, unlike the covolume approach \cite{mohamed2016discrete}.
Due to its intrinsic coordinate independent nature, DEC based implementations work on both planar domains and curved surfaces without any change. Its
independence of the embedding space dimension  as well as the
details of the embedding characterizes a key difference between DEC and the covolume method. In the latter, face normal vectors, with
which dot product of the momentum equation is performed for the discretization,
may not be unique for a simplicial mesh approximation of a curved
surface.

Here, we extend an existing stream function formulation DEC scheme \cite{mohamed2016discrete},
to the primitive variable
formulation. To compute the aerodynamic forces for the flows past
bluff / streamlined bodies, the stream function formulation requires
{\it a posteriori} computation of the pressure field from the stream function
field, which may introduce some error. The primitive variable formulation
computes the pressure field as a part of the DEC solution and is more
relevant for such physical applications. Moreover, the method is extended
to include the Coriolis force, in order to enhance its applicability
to vortical flows on rotating surfaces which are usually used to model
 planetary flows. In addition, an energy-preserving time integrator
\cite{mullen2009energy} is incorporated, to enhance applicability
of the method to study the late time behavior of such (vortical flows
on rotating surfaces) initial value problems.

\section{Primitive Variable DEC Formulation}

A primitive variable DEC formulation is presented here, extending
an existing stream function method \cite{mohamed2016discrete}. First,
the Navier-Stokes equations are expressed in exterior calculus notation, and then the DEC discretization is derived. The resultant degrees of freedom
are the primitive variables, namely the velocity 1-form and the pressure 0-form. We recommend for readers unfamiliar with DEC to read through the appendix for a primer on discrete exterior calculus before reading through
this section.

\subsection{Exterior Calculus Expression of the Navier-Stokes Equations}

For an incompressible flow of a homogeneous fluid with unit density, on a compact smooth Riemannian surface,
the vector calculus notation of Navier-Stokes equations in a rotating frame of reference are as follows. 

\begin{equation}
\frac{\partial\mathbf{v}}{\partial t}-\mu\left[-\Delta^{dR}\mathbf{v}+2\kappa\mathbf{v}\right]+\nabla_{\mathbf{v}}\mathbf{v}+\textrm{grad}_{S}p +f\hat{k}\times\mathbf{v} =0,\label{eq:1-1}
\end{equation}

\begin{equation}
\textrm{div}_{S}\mathbf{v}=0,\label{eq:2-1}
\end{equation}
where $\mathbf{v}$ is the tangential surface velocity, $p$ is the
effective surface pressure (which includes the centrifugal force), $\mu$ is the dynamic viscosity, $\Delta^{dR}$
is the surface Laplace-DeRham operator, $\kappa$ is the surface Gaussian
curvature, $\nabla_{\mathbf{v}}$ is the covariant directional derivative,
$\textrm{grad}_{S}$ is the surface gradient, $\textrm{div}_{S}$
is the surface divergence, $f$ is the Coriolis parameter, and $\hat{k}$ is the unit vector in the direction of the axis of rotation. The Laplace-DeRham operator and the Gaussian
curvature term are a result of the divergence of the deformation tensor
and the non-commutativity of the second covariant derivative in curved
spaces \cite{nitschke2017discrete}. Following \citet{mohamed2016discrete}, with the additional curvature term $-2\mu\kappa u$ and Coriolis force $f\hat{k}\times\mathbf{v}$, the 2D coordinate invariant form
of equations (\ref{eq:1-1}) - (\ref{eq:2-1}), for constant $\kappa$ (for convenience, its variation is addressed later in Sec. \ref{subsec:Discrete-exterior-calculus}), read 

\begin{equation}
\frac{\partial u }{\partial t}-\mu\ast\textrm{d}\ast\textrm{d} u -2\mu\kappa u +\ast\left( u \wedge\ast\textrm{d} u \right)+\textrm{d}p^{d} + \ast\left( u \wedge\ast f_{dual2}\right)=0.\label{eq:8}
\end{equation}

\begin{equation}
\ast\textrm{d}\ast u =0.\label{eq10}
\end{equation}

Here, $ u $ is the 1-form velocity, $\textrm{d}$ is the exterior
derivative, $\ast$ is the Hodge star operator,   
 $\wedge$ is the wedge product operator,  
  $p^{d}=p+\frac{1}{2}\left( u \left(\mathbf{v}\right)\right)$ is the effective dynamic pressure 0-form, $N$ is the space dimension, and the Coriolis force $f\hat{k}\times\mathbf{v}=\ast\left( u \wedge\ast f_{dual2}\right)$ \cite{bauer2013toward}, where $f_{dual2}$ is the dual 2-form corresponding to $f$.
Equations (\ref{eq:8}) - (\ref{eq10}) correspond to smooth exterior calculus
expressions, and are still continuous.

\subsection{Discrete Exterior Calculus Expression}

The discretization consists of the domain discretization, and discretization
of the governing equations, i.e., discrete exterior calculus expression
of the governing equations (equations (\ref{eq:8}) and (\ref{eq10})). 

\subsubsection{Domain discretization}

Two-dimensional (2D) domain discretization is considered assuming a primal simplicial
mesh. The objects of this simplicial mesh are the triangles (primal
2-simplices), edges of the triangles (primal 1-simplices), and the
vertices / nodes (primal 0-simplices). A circumcentric dual mesh
is considered, which consists of dual polygons, the edges of the dual
polygons, and the dual vertices / nodes. The dual to a primal triangle
is its circumcenter (which is a dual node), that to a primal edge
is the perpendicular dual edge connecting the circumcenters of the
two triangles sharing this primal edge, and that to a primal node
is the polygon formed by the edges dual to the primal edges connected
to this node. For a triangular mesh representing a curved surface,
the dual edges are kinked lines and the dual cells are non-planar.
Counterclockwise orientation of both the primal mesh (triangles/primal
2-simplices) and dual mesh (polygons) is assumed to be positive. The
choice of the orientation of the primal edges (primal 1-simplices)
is arbitrary, however it induces the dual edges orientations. The
orientation of the dual edges is derived simply by rotating the primal
edge orientation 90 degrees along the triangles orientation (i.e.,
counterclockwise). Moreover, there is no requirement such as the triangles need to be Delaunay \cite{doi:10.1080/15502287.2018.1446196}.  

\subsubsection{Discrete exterior calculus expression of the governing equations\label{subsec:Discrete-exterior-calculus}}

First we need to specify the correspondence between the discrete variables (the
forms) and the mesh objects (vertices, edges etc.). This requires that we define the discrete
forms, or discretize the smooth forms. We then determine the discrete
operators and substitute the smooth exterior calculus operators with
their discrete counterparts. For the current discretization, we define the velocity 1-form
$ u $ in all terms (except one in the nonlinear term which will be discussed later) on dual
edges in equation (\ref{eq:8}), i.e., we chose the velocity 1-form
$ u $ to be a dual 1-form. The dual 1-form represents the
velocity integration along the dual edges and hence corresponds to
the mass flux across the primal edges. With this choice, for consistency,
each term (time derivative, convection, diffusion, pressure gradient)
in equation (\ref{eq:8}) must be a dual 1-form. This choice of the
velocity 1-form requires the dynamic pressure 0-form $p^{d}$ to be
defined at the dual nodes or the triangles circumcenters (to be defined
as a dual 0-form), so that the pressure gradient term $\textrm{d}p^{d}$
is a dual 1-form, consistent with the choice of $ u $ as the
dual 1-form. For consistency, since the convective term $\ast\left( u \wedge\ast\textrm{d} u \right)$
is a dual 1-form, the wedge product term $\left( u \wedge\ast\textrm{d} u \right)$
is a primal 1-form (defined on primal edges). Since $ u $
is a 1-form, and hence $\ast\textrm{d} u $ is a 0-form (in
2D), and therefore the operands of the wedge product are a 1-form
and a 0-form, such that the resultant (primal 1-form) must be defined
on primal edges. This requires that both wedge product operands are
defined on primal mesh objects. Accordingly, the velocity 1-form in
the first operand $\left( u \right)$ of the wedge product
term is defined on the primal edges, whereas the velocity 1-form in
the second operand $\left(\ast\textrm{d} u \right)$ is defined
on the dual edges, so that the 0-form $\ast\textrm{d} u $
is defined on the primal nodes. Following \citet{mohamed2016discrete}, the definitions and expressions for the discrete
$\textrm{d}$, $\ast$, and $\wedge$ operators, the DEC discretization of equation (\ref{eq:8}) now reads (in 2D)

\begin{eqnarray}
&\frac{U^{n+1}-U^{n}}{\triangle t}-\mu\ast_{1}\textrm{d}_{0}\ast_{0}^{-1}\left[\left[-\textrm{d}_{0}^{T}\right]U+\textrm{d}_{b}V\right]-2\mu\kappa U \nonumber\\&+\ast_{1}W_{V}\ast_{0}^{-1}\left[\left[-\textrm{d}_{0}^{T}\right]U+\textrm{d}_{b}V\right]+\textrm{d}_{1}^{T}P^{d} + \ast_{1}W_{V}\ast_{0}^{-1}f_{dual2}=0.\label{eq:1}
\end{eqnarray}

Here, $U$ is the vector containing the discrete dual velocity 1-forms
for all mesh dual edges, $V$ is the vector containing the discrete
primal velocity 1-forms for all mesh primal edges, and $P^{d}$ is
the vector containing discrete effective dynamic pressure 0-forms for all mesh
dual vertices, $\Delta t$ is the discrete time interval between the
current time instant $n+1$ and the previous time instant $n$. Note
that the time-centering of all terms except the one involving the time derivative
is discussed later in sections \ref{subsec:Euler-first-order}
and \ref{subsec:Energy-preserving-time}. 
The matrix $W_{V}$ is defined as $W_{V}=0.5diag(V)\left|\textrm{d}_{0}\right|$
and is a sparse $N_{1}\times N_{0}$ matrix, where $N_{1}$ is the
number of primal edges and $N_{0}$ is the number of primal nodes.
Moreover, the term $\ast_{0}^{-1}\left[-\textrm{d}_{0}^{T}\right]U$
represents vorticity and is a vector with $N_{0}$ entries. The term
$W_{V}\ast_{0}^{-1}\left[-\textrm{d}_{0}^{T}\right]U$ is equivalent
to taking the average, at each primal edge, of the vorticity $\ast_{0}^{-1}\left[-\textrm{d}_{0}^{T}\right]U$
evaluated at the edge's nodes and multiply it by the edge\textquoteright s
tangential velocity 1-form $v$. Thus, this term evaluates the first wedge
product. Moreover, since
$\ast f_{dual2}$ is a primal 0-form, the 1-form velocity $ u $ in
the second wedge product also has to be the primal 1-form for consistency. Thus the term $\ast_{1}W_{V}\ast_{0}^{-1}f_{dual2}$ evaluates the second wedge product. The tangential velocity 1-form $v$ at each primal edge is
evaluated following \citet{mohamed2016discrete}.

The boundary of the dual 2-cells to the primal nodes on the
domain boundary includes primal boundary edges. Since $\left[-\textrm{d}_{0}^{T}\right]$
matrix is the boundary operator of these dual 2-cells, it is complemented
by an additional operator $\textrm{d}_{b}$ to account for the primal
boundary edges. The operation $\left[-\textrm{d}_{0}^{T}\right]U$
evaluates the circulation of the velocity 1-forms $ u $ along
the dual 2-cells boundaries. The operation $\textrm{d}_{b}V$ complements
this circulation for the dual 2-cells part of whose boundaries consists
of primal edges. The matrix $\textrm{d}_{b}=0.5\left|\textrm{d}_{0}^{T}\right|\textrm{diag}\left(\textrm{d}_{1}^{T}\mathds{1}\right)$
, where $\left|\textrm{d}_{0}^{T}\right|$ is the matrix $\textrm{d}_{0}^{T}$
with all entries made positive and $\mathds{1}$ is a vector of ones
with $N_{2}$ (number of primal 2-simplices / triangles) entries,
and $\textrm{diag}\left(\cdot\right)$ is a diagonal matrix consisting
of the enclosed vector entries. 

Since the mass flux primal 1-form (the mass flux across the primal
edge) is defined as the integral of the normal velocity (normal to
the primal edge) along the primal edge, the vector containing mass
flux primal 1-form $u^{\ast}$ for all mesh primal edges can be expressed
as $U^{*}=-\ast_{1}^{-1}U$ or $U=\ast_{1}U^{*}$. After this substitution,
Eq. (\ref{eq:1}) reads 

\begin{eqnarray}
&\ast_{1}\left[\frac{\left(U^{*}\right)^{n+1}-\left(U^{*}\right)^{n}}{\triangle t}\right]-\mu\ast_{1}\textrm{d}_{0}\ast_{0}^{-1}\left[\left[-\textrm{d}_{0}^{T}\right]\ast_{1}U^{*}+\textrm{d}_{b}V\right]-2\mu\kappa\ast_{1}U^{\ast} \nonumber\\&+\ast_{1}W_{V}\ast_{0}^{-1}\left[\left[-\textrm{d}_{0}^{T}\right]\ast_{1}U^{*}+\textrm{d}_{b}V\right]+\textrm{d}_{1}^{T}P^{d} + \ast_{1}W_{V}\ast_{0}^{-1}f_{dual2}=0.\label{eq:2-2}
\end{eqnarray}

Applying the Hodge star $\ast_{1}^{-1}$ to equation (\ref{eq:2-2}),
and considering the property $\ast_{1}^{-1}\ast_{1}=-1$, we have

\begin{eqnarray}
&-\frac{\left(U^{*}\right)^{n+1}-\left(U^{*}\right)^{n}}{\triangle t}+\mu\textrm{d}_{0}\ast_{0}^{-1}\left[\left[-\textrm{d}_{0}^{T}\right]\ast_{1}U^{*}+\textrm{d}_{b}V\right]+2\mu\kappa U^{*} \nonumber\\& -W_{V}\ast_{0}^{-1}\left[\left[-\textrm{d}_{0}^{T}\right]\ast_{1}U^{*}+\textrm{d}_{b}V\right]+\ast_{1}^{-1}\textrm{d}_{1}^{T}P^{d} - W_{V}\ast_{0}^{-1}f_{dual2}=0.\label{eq:2}
\end{eqnarray}

In general, $\kappa$ may not be uniform. For domains with non-uniform $\kappa$, it may be treated as a primal 0-form (defined at the primal vertices). Now the curvature term $\left(2\mu\kappa U^{\ast}\right)$ is modified and Eq. (\ref{eq:2}) reads

\begin{eqnarray}
&-\frac{\left(U^{*}\right)^{n+1}-\left(U^{*}\right)^{n}}{\triangle t}+\mu\textrm{d}_{0}\ast_{0}^{-1}\left[\left[-\textrm{d}_{0}^{T}\right]\ast_{1}U^{*}+\textrm{d}_{b}V\right]+2\mu K U^{*} \nonumber\\& -W_{V}\ast_{0}^{-1}\left[\left[-\textrm{d}_{0}^{T}\right]\ast_{1}U^{*}+\textrm{d}_{b}V\right]+\ast_{1}^{-1}\textrm{d}_{1}^{T}P^{d} - W_{V}\ast_{0}^{-1}f_{dual2}=0.\label{eq:2_a}
\end{eqnarray}

Here,  $K$ is half times a matrix containing values of $\kappa$ at the nodes of each edge, and it  accounts for the wedge product of the primal zero form $\kappa$ with the primal 1-form $u^{\ast}$. With reference to temporal discretization, we first implemented
the Euler first order time integration scheme. Later,  in order
to enhance the ability to study the late-time behavior of initial
value problems, we implemented the energy-preserving time integration
\cite{mullen2009energy}. Both time integration schemes are discussed below.

\subsubsection{Euler first order time integration\label{subsec:Euler-first-order}}

We treat the viscous, curvature and pressure gradient terms implicitly, and the convection and Coriolis force
terms explicitly. In order to obtain a linear representation for the
convective term, we evaluate the tangential velocity 1-forms $V$
through an interpolation of previously-known normal velocity 1-forms
$ U $. Therefore, we have

\begin{eqnarray}
&-\frac{\left(U^{*}\right)^{n+1}-\left(U^{*}\right)^{n}}{\triangle t}+\mu\textrm{d}_{0}\ast_{0}^{-1}\left[\left[-\textrm{d}_{0}^{T}\right]\ast_{1}\left(U^{*}\right)^{n+1}+\textrm{d}_{b}\left(V\right)^{n}\right]+2\mu K \left(U^{\ast}\right)^{n+1} \nonumber\\& -\left(W_{V}\right)^{n}\ast_{0}^{-1}\left[\left[-\textrm{d}_{0}^{T}\right]\ast_{1}\left(U^{*}\right)^{n}+\textrm{d}_{b}\left(V\right)^{n}\right]+\ast_{1}^{-1}\textrm{d}_{1}^{T}\left(P^{d}\right)^{n+1} \nonumber\\& - \left(W_{V}\right)^{n}\ast_{0}^{-1}f_{dual2}=0,\label{eq:2-3}
\end{eqnarray}
or

\begin{equation}
\left[-\frac{1}{\triangle t} I +\mu\textrm{d}_{0}\ast_{0}^{-1}\left[-\textrm{d}_{0}^{T}\right]\ast_{1} +2\mu K \right]\left(U^{*}\right)^{n+1}+\ast_{1}^{-1}\textrm{d}_{1}^{T}\left(P^{d}\right)^{n+1}=F,\label{eq:3}
\end{equation}
where $I$ is an identity matrix, and
\begin{eqnarray}
&F=-\frac{1}{\triangle t}\left(U^{*}\right)^{n}-\mu\textrm{d}_{0}\ast_{0}^{-1}\textrm{d}_{b}\left(V\right)^{n}\nonumber\\&+\left(W_{V}\right)^{n}\ast_{0}^{-1}\left[\left[-\textrm{d}_{0}^{T}\right]\ast_{1}\left(U^{*}\right)^{n}+\textrm{d}_{b}\left(V\right)^{n} + f_{dual2}\right].\label{eq:4}
\end{eqnarray}

The continuity equation is expressed as

\begin{equation}
\left[\textrm{d}_{1}\right]\left(U^{*}\right)^{n+1}+\left[0\right]\left(P^{d}\right)^{n+1}=0.\label{eq:5}
\end{equation}

Equations (\ref{eq:3}) - (\ref{eq:5}) represent a set of linear
equations with $U^{\ast}$ and $P^{d}$ as the solution vectors, which
is solved along with usual boundary conditions depending on the considered test case. 

\subsubsection{Energy-preserving time integration\label{subsec:Energy-preserving-time}}

The midpoint integration is assumed for the update in time of the advection, diffusion,
curvature, pressure gradient and Coriolis force terms \cite{mullen2009energy},
i.e., these terms are evaluated at time $n+\frac{1}{2}$. Thus we
have,

\begin{eqnarray}
%\begin{aligned}
&-\frac{\left(U^{*}\right)^{n+1}-\left(U^{*}\right)^{n}}{\triangle t}+\frac{1}{2}\left[\mu\textrm{d}_{0}\ast_{0}^{-1}\left(\left[-\textrm{d}_{0}^{T}\right]\ast_{1}\left(U^{*}\right)^{n+1}+\textrm{d}_{b}\left(V\right)^{n+1}\right)  \right. \nonumber\\ & \left. +\mu\textrm{d}_{0}\ast_{0}^{-1}\left(\left[-\textrm{d}_{0}^{T}\right]\ast_{1}\left(U^{*}\right)^{n}+\textrm{d}_{b}\left(V\right)^{n}\right)\right]+\frac{1}{2}\left[2\mu K \left(U^{\ast}\right)^{n+1}+2\mu K \left(U^{\ast}\right)^{n}\right] \nonumber\\& -\frac{1}{2}\left[\left(W_{V}\right)^{n+1}\ast_{0}^{-1}\left(\left[-\textrm{d}_{0}^{T}\right]\ast_{1}\left(U^{*}\right)^{n+1}+\textrm{d}_{b}\left(V\right)^{n+1}\right) \right. \nonumber\\& \left. +\left(W_{V}\right)^{n}\ast_{0}^{-1}\left(\left[-\textrm{d}_{0}^{T}\right]\ast_{1}\left(U^{*}\right)^{n}+\textrm{d}_{b}\left(V\right)^{n}\right)\right]+\ast_{1}^{-1}\textrm{d}_{1}^{T}\left(P^{d}\right)^{n+\frac{1}{2}} \nonumber\\& - \frac{1}{2}\left[\left(W_{V}\right)^{n+1}\ast_{0}^{-1}f_{dual2} + \left(W_{V}\right)^{n}\ast_{0}^{-1}f_{dual2} \right]=0.
%\end{aligned}
\end{eqnarray}

Rearranging the terms, we have

\begin{eqnarray}
&\left[-\frac{1}{\Delta t} I +\mu K +\frac{1}{2}\mu\textrm{d}_{0}\ast_{0}^{-1}\left[-\textrm{d}_{0}^{T}\right]\ast_{1}-\frac{1}{2}\left(W_{V}\right)^{n+1}\ast_{0}^{-1}\left[-\textrm{d}_{0}^{T}\right]\ast_{1}\right]\left(U^{*}\right)^{n+1} \nonumber\\&+\ast_{1}^{-1}\textrm{d}_{1}^{T}\left(P^{d}\right)^{n+\frac{1}{2}}=F,\label{eq:energy_preserving}
\end{eqnarray}
with

\begin{eqnarray}
&F=-\frac{1}{\Delta t} \left(U^{*}\right)^{n} -\mu K \left(U^{*}\right)^{n}-\frac{1}{2}\mu\textrm{d}_{0}\ast_{0}^{-1}\left(\left[-\textrm{d}_{0}^{T}\right]\ast_{1}\left(U^{*}\right)^{n}+\textrm{d}_{b}\left(V\right)^{n} \right. \nonumber\\ & \left. +\textrm{d}_{b}\left(V\right)^{n+1}\right) +\frac{1}{2}\left(W_{V}\right)^{n+1}\ast_{0}^{-1}\textrm{d}_{b}\left(V\right)^{n+1} \nonumber\\& +\frac{1}{2}\left(W_{V}\right)^{n}\ast_{0}^{-1}\left(\left[-\textrm{d}_{0}^{T}\right]\ast_{1}\left(U^{*}\right)^{n}+\textrm{d}_{b}\left(V\right)^{n}\right) \nonumber\\& + \frac{1}{2}\left[\left(W_{V}\right)^{n+1} + \left(W_{V}\right)^{n} \right] \ast_{0}^{-1}f_{dual2}.
\end{eqnarray}

Equation (\ref{eq:energy_preserving}) along with equation (\ref{eq:5}),
and the relevant boundary conditions (depending on the test case) form
a non-linear system of equations. These are solved using Picard's iterative
method, with previous time-step value used as an initial guess for
each time-step and a convergence criterion of $L_{2}$ norm on the residual
of $\left\Vert R\right\Vert _{2}\leq10^{-8}$. It is worth noting that the current discretization is different from that in \cite{mullen2009energy} for the convective term spatial discretization and the inclusion of both surface curvature effect and Coriolis force.

\section{Results and Discussions}

In order to assess the accuracy of the scheme, several test cases are
performed. All test cases of flow past a cylinder are solved using the Euler time integration scheme. The energy-preserving time integration scheme is then employed for
the remaining test cases. 

\subsection{Flow Past a Stationary Circular Cylinder\label{subsec:Flow-Past-a}}

The computational domain consists of a rectangular plane with a circular
cylinder embedded on it. The circular cylinder contour embedded on
the surface is generated by intersecting a cylinder of diameter D
= 1 with the surface. The domain lengths upstream and downstream of
the cylinder are 10D and 60D, respectively, and the domain width on
either side of the cylinder is 10D. These domain sizes are usual in
computational investigations of flow past a cylinder. Uniform normal
velocity of unity is assumed at the inlet. Outflow boundary condition
is assumed at the outlet. Free-slip boundary condition is considered on the top and bottom boundaries, whereas no-slip boundary condition is assumed on the cylinder wall. The computational domain is discretized using a triangular mesh
(consisting of approximately 160k elements) with locally refined mesh in the vicinity of the
cylinder and in the wake. Reynolds numbers ($Re=\rho u_{\infty}D/\mu$) = 40, 100, and 1000 are considered. Here, $\rho$ and $u_{\infty}$ are the fluid density (taken to be unity) and the free stream velocity at the inlet, respectively.

The flow exerts pressure and viscous forces on the cylinder, and the
net force acting on the cylinder is expressed as \cite{Shi2004} 
\begin{equation}
\vec{F}=\int_{S}-p\vec{n}dA+\mu\int_{S}\vec{\omega}\times\vec{n}dA,
\end{equation}
where $p$ is the static pressure, $\vec{n}$ is the unit normal facing
outward of the cylinder, and $\vec{\omega}$ is the vorticity vector
(normal to the embedding surface). The integration is carried out
over the total surface area $S$ (perimeter here) of the cylinder.
The discrete expression of the force vector reads

\begin{equation}
\vec{F}=\sum_{k=1}^{N}-p_{k}\vec{n}_{k}l_{k}+\mu\sum_{k=1}^{N}l_{k}\vec{\omega}_{k}\times\vec{n}_{k},\label{eq:9}
\end{equation}
where the summation is over the cylinder $N$ boundary edges. Here, global $(x,y)$ coordinate system
is considered with the cylinder axis intersection with the embedding
surface as the origin. The $x$ and $y$ coordinates are along the
streamwise and cross-streamwise directions, respectively. The static
pressure $p_{k}$ is in the triangles adjacent to the boundary edges,
$\vec{n}_{k}$ is the unit vector normal to the edge, and $l_{k}$
is the edge length. The vorticity vector is $\vec{\omega}_{k}=\vec{n}_{k,s}|\omega_{k}|$,
where $\vec{n}_{k,s}$ is the unit normal to the embedding surface
at the boundary edge midpoint. The vorticity magnitude $|\omega_{k}|$
is calculated by averaging the known vorticity on both of the edge
nodes. The drag force $F_{d}$ and the lift force $F_{l}$ are respectively
the streamwise and cross-streamwise components of the force vector
$\vec{F}$, i.e., are the $x$ and $y$ components of the force vector
$\vec{F}$. We calculate the drag coefficient, lift coefficient, and
Strouhal number as $C_{d}=F_{d}/\left(\frac{1}{2}\rho u_{\infty}^{2}D\right)$,
$C_{l}=F_{l}/\left(\frac{1}{2}\rho u_{\infty}^{2}D\right)$, and $St=fD/u_{\infty}$,
respectively. Here, $\rho$ is the fluid density (taken to be unity),
$u_{\infty}$ is the flow velocity at the inlet boundary of the domain
(also taken to be unity), and $f$ is the vortex shedding frequency. 

Table \ref{table1} reports the drag coefficient mean values. The
computed values are $1.605$, $1.386$, and $1.544$ at $Re=40,100$,
and $1000$, respectively. They agree well with that reported in the
literature (e.g., at $Re=40$: $1.6$ \cite{russell2003cartesian},
$1.618$ \cite{kawaguti1953numerical}; at $Re=100$: $1.38$ \cite{russell2003cartesian},
$1.37$ \cite{le2006immersed}; at $Re=1000$: $1.5144$ \cite{henderson1995unstructured},
$1.5$ \cite{sheard2005computations}). Table \ref{table1} also
reports the RMS values of the lift coefficient. We compute the values
to be $0.248$, and $1.081$ at $Re=100$, and $1000$, respectively.
They are within the range of values reported in the literature (e.g.,
at $Re=100$: $0.233$ \cite{Stalberg2006}; at $Re=1000$: $1.0494$
\cite{henderson1995unstructured}). In passing, it is worth noting
that the predicted values of the mean drag coefficient and RMS lift
coefficient using the stream function formulation \cite{mohamed2016discrete}
are $1.337$ and $0.171$, respectively at $Re=100$. The drag coefficient
is predicted correctly, but the lift coefficient is under-predicted.
The under-prediction of the lift coefficient may be, as already discussed
before (see section \ref{sec:Introduction}), because of the error
induced in posterior computation of the pressure field. Table~\ref{table1}
reports the Strouhal number values. The Strouhal number is computed
as $0.171$, and $0.239$ at $Re=100$, and $1000$, respectively,
which is within the range of values reported in the literature (e.g.,
at $Re=100$: $0.172$ \cite{russell2003cartesian}, $0.166$ \cite{Stalberg2006};
at $Re=1000$: $0.237$ \cite{henderson1995unstructured}, $0.244$
\cite{rajani2005laminar}).

\begin{table*}
\centering{}\centering{}%
\begin{tabular}{cccccccc}
\hline 
Reference  & $Re=40$  & \multicolumn{3}{c}{$Re=100$} & \multicolumn{3}{c}{$Re=1000$}\tabularnewline
 & $C_{d}$  & $C_{d}$  & $C_{l}$  & $St$  & $C_{d}$  & $C_{l}$  & $St$\tabularnewline
 &  & (mean) & (RMS) &  & (mean) & (RMS) & \tabularnewline
\hline 
Present study  & 1.605  & 1.386  & 0.248  & 0.171  & 1.544  & 1.081  & 0.239\tabularnewline
\citet{russell2003cartesian}  & 1.600  & 1.380  & -  & 0.172  & -  & -  & -\tabularnewline
\citet{le2006immersed}  & 1.560  & 1.370  & -  & 0.160  & -  & -  & -\tabularnewline
\citet{Stalberg2006}  & 1.53  & 1.32  & 0.233  & 0.166  & -  & -  & -\tabularnewline
\citet{kawaguti1953numerical}  & 1.618  & -  & -  & -  & -  & -  & -\tabularnewline
\citet{Dennis1970}  & 1.522  & -  & -  & -  & -  & -  & -\tabularnewline
\citet{henderson1995unstructured}  & -  & -  & -  & -  & 1.5144  & 1.0494  & 0.237 \tabularnewline
\citet{sheard2005computations}  & -  & -  & -  & -  & 1.500  & -  & -\tabularnewline
\citet{rajani2005laminar}  & -  & 1.342  & -  & 0.166  & 1.492  & -  & 0.244 \tabularnewline
\hline 
\end{tabular}\caption{\label{table1}Mean drag coefficient, RMS lift coefficient, and Strouhal
number for flow past a stationary circular cylinder case.}
\end{table*}

\begin{figure}
\centering{}\includegraphics[scale=0.4]{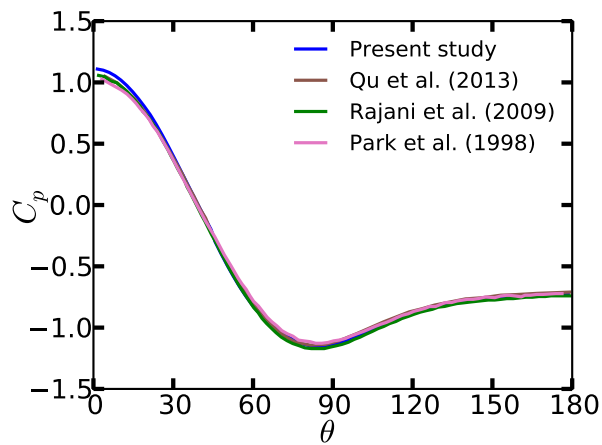}\caption{Pressure coefficient as a function of angular position on the cylinder
for flow past a stationary circular cylinder case. \label{fig:Pressure-coefficinet-as}}
\end{figure}

Figure \ref{fig:Pressure-coefficinet-as} shows the plots of time
averaged pressure coefficient $C_{p}=\left(p-p_{\infty}\right)/ \\ \left(\frac{1}{2}\rho u_{\infty}^{2}\right)$
as a function of the angular position on the cylinder surface ($\theta$)
at $Re=100$. Here, $p_{\infty}$ is the free stream pressure. The current results agree well with literature. 

The instantaneous flow separation angle values $\left(\theta_{sep}\right)$
are summarized in table~\ref{table2}. They are determined to be
$54.75^{\circ}$, $116.55$, $108.87$ at $Re=40$, $100$ and $1000$,
respectively. They are within the range mentioned in the literature
(e.g., \cite{kawaguti1966numerical}, \cite{taneda1956experimental},
\cite{rajani2005laminar}). 

\begin{table*}
\centering{}\centering{}%
\begin{tabular}{cccc}
\hline 
Reference & $Re=40$ (from  & $Re=100$ (from  & $Re=1000$ (from \tabularnewline
 & downstream & upstream & upstream\tabularnewline
 & stagnation point)  & stagnation point)  & stagnation point)\tabularnewline
\hline 
Present study  & 54.75  & 116.55  & 108.87\tabularnewline
\citet{kawaguti1966numerical}  & 53.7  & -  & -\tabularnewline
\citet{taneda1956experimental}  & 53.0  & -  & -\tabularnewline
\citet{rajani2005laminar}  & -  & 117.05 $\pm$ 2.83  & 100.17 $\pm$ 18.9 \tabularnewline
 &  & (periodic variation)  & (periodic variation)\tabularnewline
\hline 
\end{tabular}\caption{\label{table2}Values of flow separation angle (in degrees) for flow
past a stationary circular cylinder case.}
\end{table*}

There is an attached stationary pair of vortices in the cylinder wake
at $Re=40$, the length of which is defined as the distance between
the cylinder center and the end of the stationary vortex. The vortex
pair length is determined to be $2.78$, which is within the range
reported in the literature (e.g., $3.01$ \cite{kawaguti1966numerical},
and $2.64$ \cite{apelt1961steady}). 

\subsection{Flow Past a Rotating Circular Cylinder}

The same flow configuration as described in section in \ref{subsec:Flow-Past-a}
is used, except that the cylinder now has a non-zero tangential velocity.
Reynolds number of $200$ is considered. Non-dimensional tangential
velocities $\alpha=\left(D/2\right)\omega/u_{\infty}=2.5$ and $\alpha=3.0$
are employed. Here, $\omega$ is the tangential velocity of the cylinder.
Since the cylinder is rotating, there is non-zero lift force acting
on the cylinder. The lift coefficient $\left(C_{L}\right)$ (as already
defined in section \ref{subsec:Flow-Past-a}) values are reported
in table \ref{tab:Lift-coefficient-for}. For $\alpha=2.5$ and $3$,
the values of $C_{L}$ are found to be $-7.641$ and $-10.413$, respectively,
which agree with the values $-7.63$ and $-10.329$ reported in \cite{MITTAL2003}.

\begin{table}
\centering{}%
\begin{tabular}{|c|c|c|}
\hline 
Reference & $\alpha$ & $C_{L}$\tabularnewline
\hline 
\hline 
Present study & $2.5$ & $-7.641$\tabularnewline
\cline{2-3} 
 & $3.0$ & $-10.413$\tabularnewline
\hline 
\citet{MITTAL2003} & $2.5$ & $-7.63$\tabularnewline
\cline{2-3} 
 & $3.0$ & $-10.329$\tabularnewline
\hline 
\end{tabular}\caption{Lift coefficient for the flow past a rotating circular cylinder case\label{tab:Lift-coefficient-for}}
\end{table}

\begin{figure}

\subfloat[]{\centering{}\includegraphics[scale=0.33]{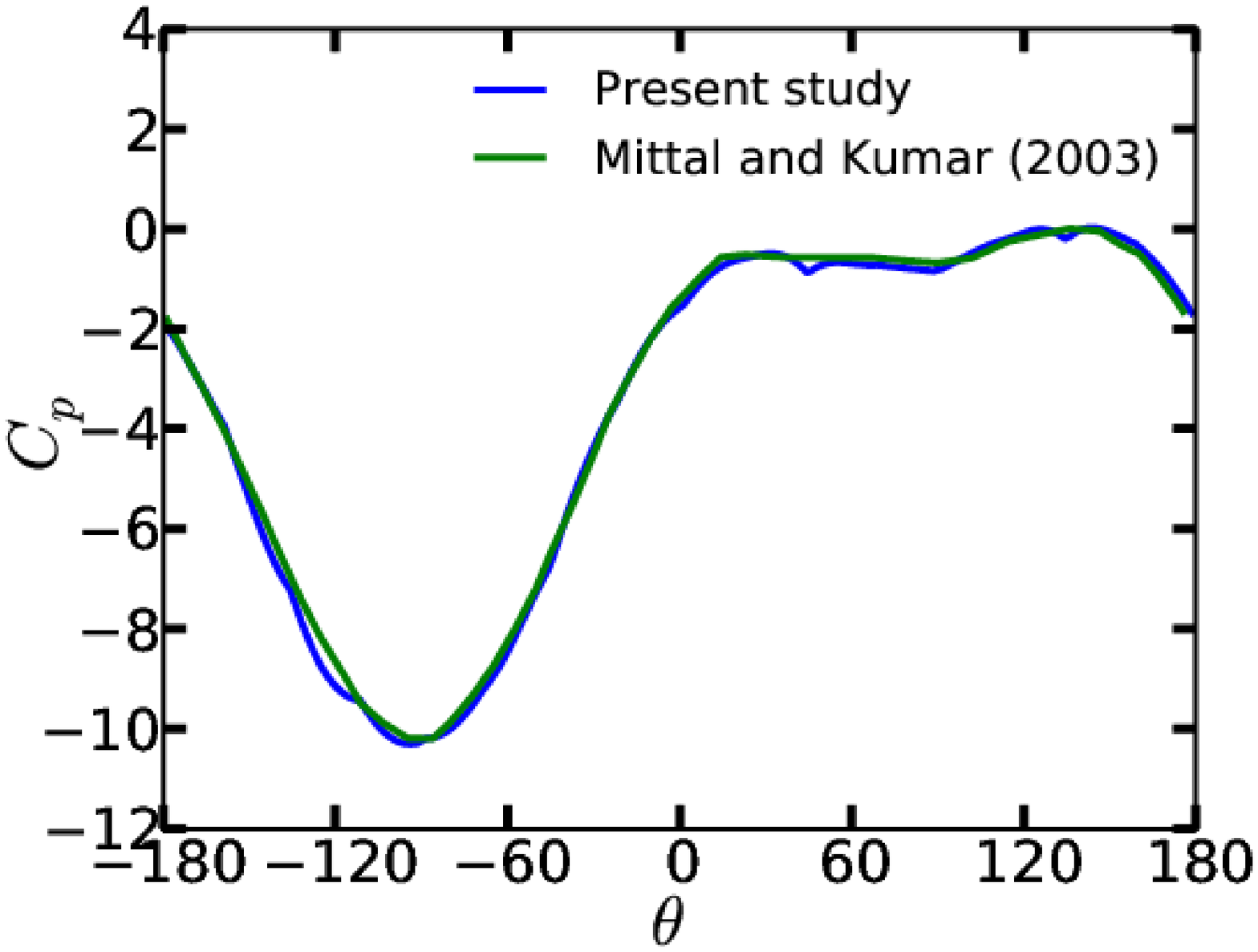}}\subfloat[]{
\begin{centering}
\includegraphics[scale=0.33]{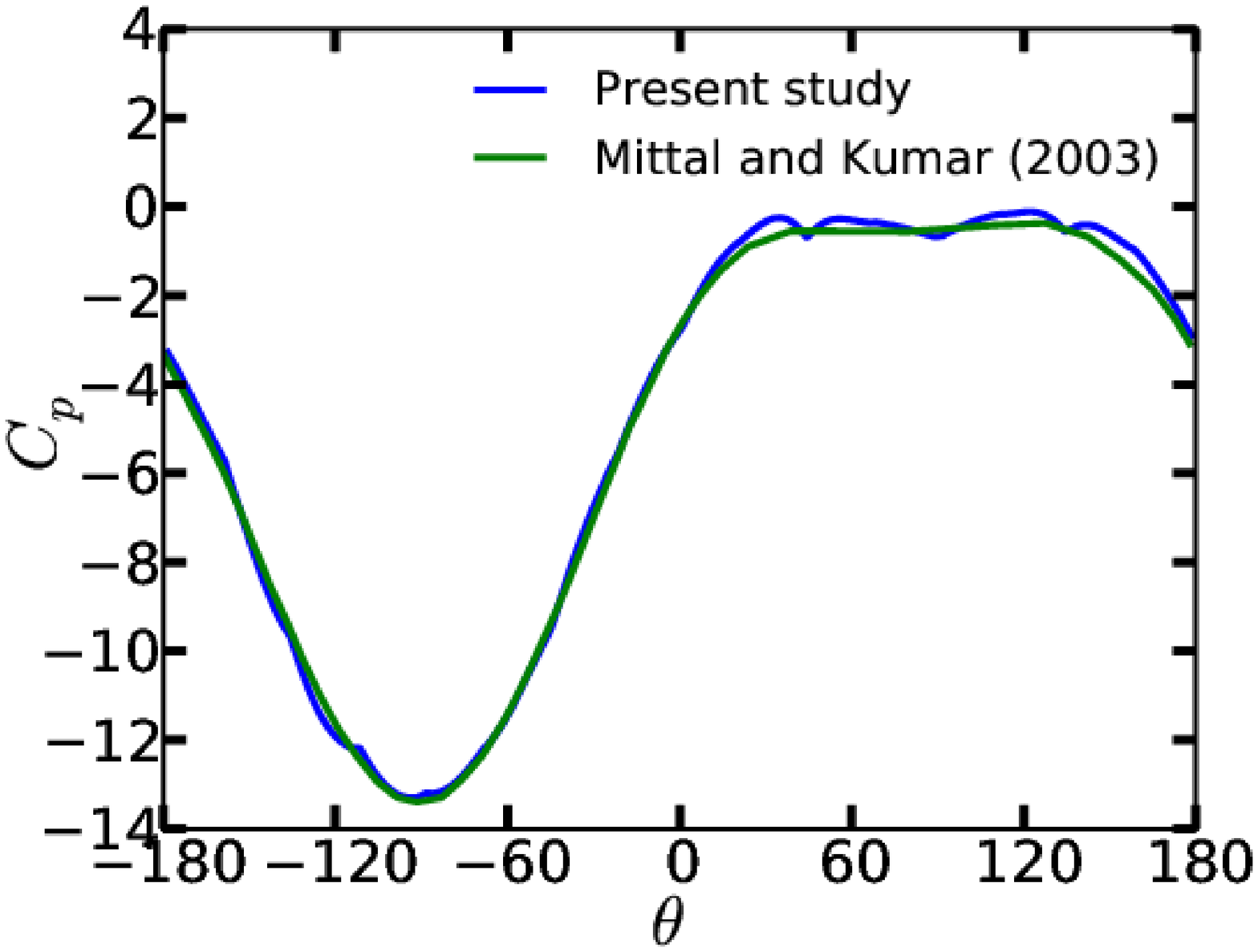}
\par\end{centering}
}\caption{Pressure coefficient as a function of angular position on the cylinder
for the flow past a rotating cylinder case. (a) $\alpha=2.5$, (b) $\alpha=3.0$. \label{fig:Pressure-coefficinet-as-1}}
\end{figure}

Figure \ref{fig:Pressure-coefficinet-as-1} shows the plots of pressure
coefficient as a function of the angular position on the cylinder
surface ($\theta$) for $\alpha=2.5$ and $3.0$ at $Re=200$. They
agree with that reported in \cite{MITTAL2003}.

\subsection{Flow Past a Stationary Circular Cylinder on a Locally Curved Surface\label{subsec:Locally-Curved}}

\begin{figure} 		
\centering 		
\includegraphics[width=0.8\columnwidth]{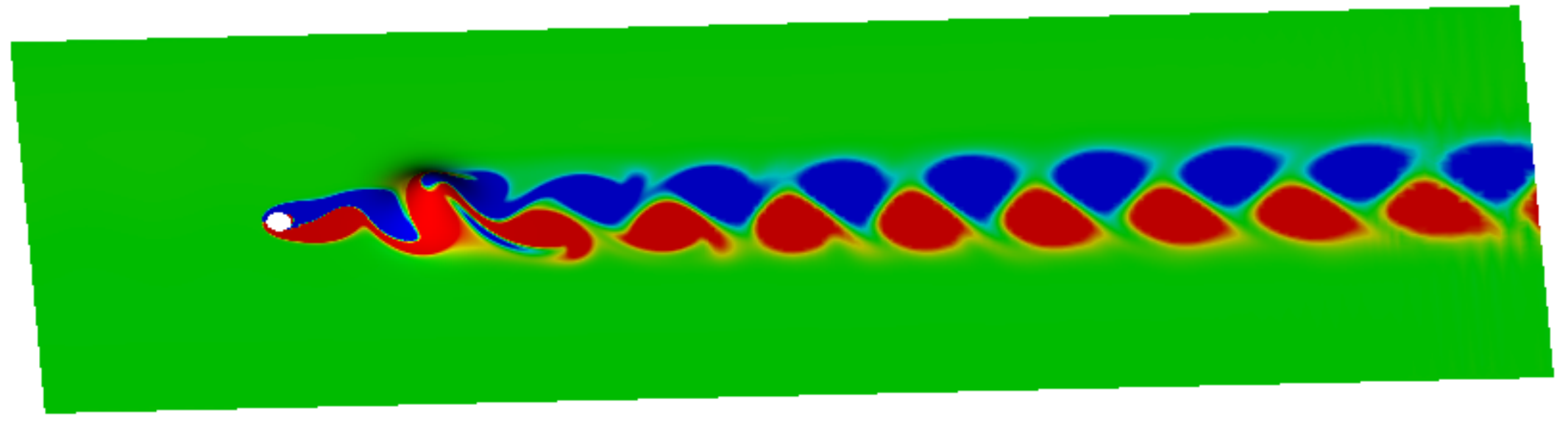} 		
\caption{Instantaneous cylinder wake for flow past a stationary cylinder with a Gaussian bump in the wake at $Re=100$.} 		
\label{gaussian_bump} 	
\end{figure}

The same flow configuration as the one described in section \ref{subsec:Flow-Past-a} is employed except that now there is a bump present in the cylinder wake. The bump is described by

\begin{equation}
z=h\exp\left(-\left(\left(x-x_{0}\right)^{2}+\left(y-y_{0}\right)^{2}\right)/\left(2r_{0}^{2}\right)\right)
\end{equation}

\noindent where the parameters describing the bump location read: $x_{0}=6.0,y_{0}=0.5,h=1.6,1/\left(2r_{0}^{2}\right)=0.4$. The global Cartesian coordinate system as described in section \ref{subsec:Flow-Past-a} is employed, with an additional coordinate $z$ to describe the bump. Figure \ref{gaussian_bump} shows an instantaneous cylinder wake at Reynolds number, based on cylinder diameter, $Re=100$. A comparison with the case without bump shows some disruption of the wake due to the vorticity generated by the bump. However, the vorticity generated at the cylinder wall dominates and the effect of the presence of bump is insignificant. The mean drag coefficient, the RMS lift coefficient, and Strouhal number are computed to be $1.37$, $0.255$, and $0.168$, respectively. They are quite close to the values reported in table \ref{table1} for the case without bump, quantitatively confirming the observed insignificant effect of the bump. 

 	%%%%%%%%%%%%%%%%% 

\subsection{Flow Past Stationary Square and Triangular Cylinders on a Spherical Surface}
\begin{figure}
\centering{}
\subfloat[]{\centering{}\includegraphics[scale=0.25]{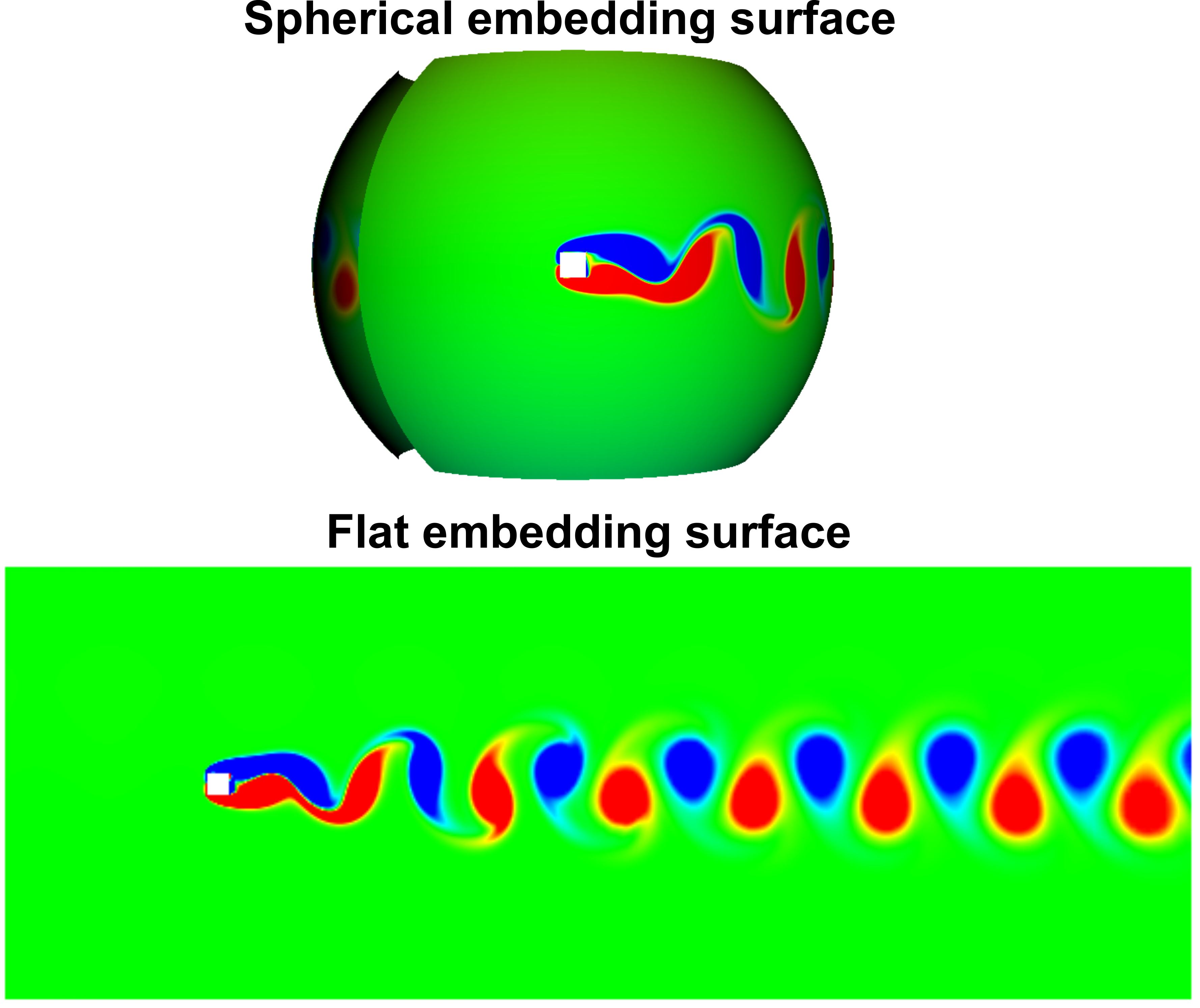}} \\ \subfloat[]{
\begin{centering}
\includegraphics[scale=0.25]{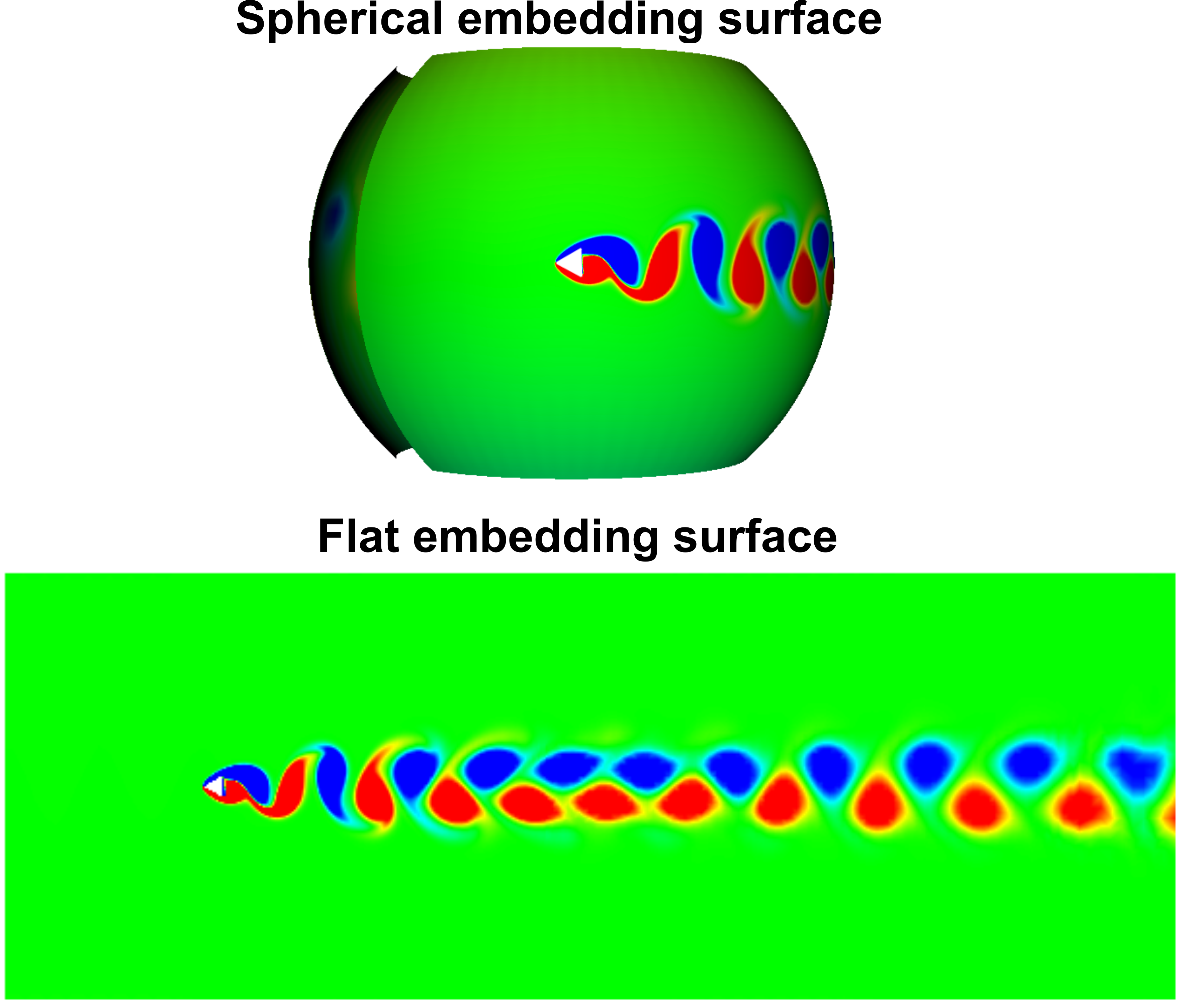}
\par\end{centering}
}\caption{Instantaneous vortex streets for flow past a cylinder on a spherical embedding surface. (a) Square cylinder, (b) Triangular cylinder. \label{fig:vortex_streets_spherical_surface}}
\end{figure}

The computational domain consists of a square or an equilateral triangular cylinder contour, both with edge length unity, embedded on a spherical surface of radius $R=12$. One of the edges of the square cylinder (contour) faces the upstream flow and the opposite edge faces the downstream flow. The vertical edge of the triangular cylinder (contour) faces the downstream flow. The domain size is similar to the one described in section \ref{subsec:Flow-Past-a}, except that now the lengths are geodesic distances. The same boundary conditions as described in section \ref{subsec:Flow-Past-a} are employed. Reynolds number, based on the edge length, $Re=100$ is considered. Figure \ref{fig:vortex_streets_spherical_surface} shows the instantaneous vortex streets, along with that for the case of flat embedding surface for a reference. Qualitatively, the vortex streets on the spherical embedding surface look similar to that on the flat embedding surface. For the square cylinder case, the mean drag coefficient, RMS lift coefficient and the Strouhal number are computed to be $1.477$, $0.2$ and $0.148$, respectively. We compare them to the corresponding values for the flat embedding surface, and the relative difference is found to be $1.2$\% and $0.67$\%, respectively for the mean drag coefficient and the Strouhal number. The RMS lift coefficient is found to be identical.  Similarly, for the triangular cylinder, the mean drag coefficient, the RMS lift coefficient and the Strouhal number are computed to be $1.77$, $0.305$ and $0.2$, respectively. The corresponding difference is found to be $2.69$\%, $4.98$\% and $1.96$\% relative to the flat embedding surface case.

\subsection{Flow Past an Airfoil}

\begin{figure} 		
\centering 		
\includegraphics[width=0.6\columnwidth]{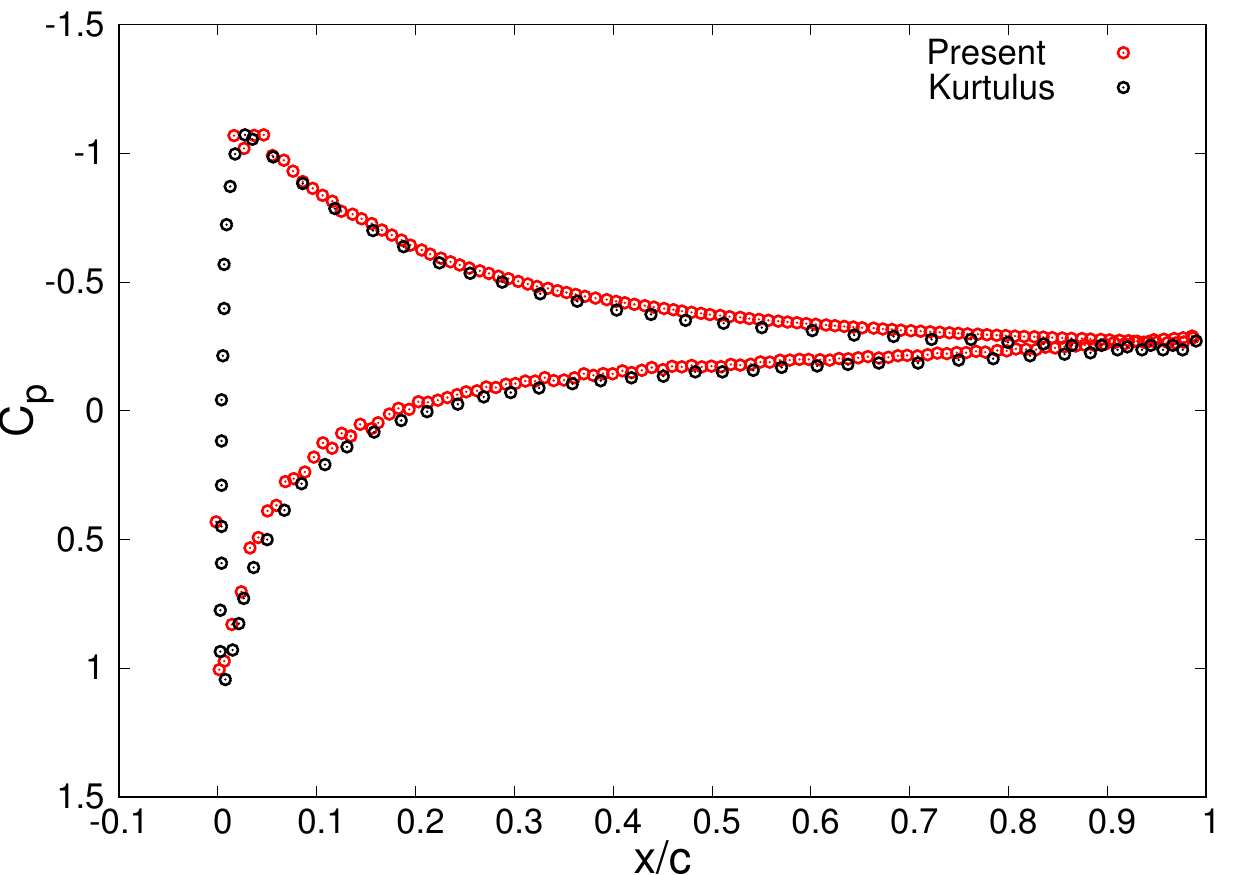} 		
\caption{Mean pressure distribution for NACA0012 at an angle of attack $\alpha = 9^o$ and $Re = 1000$.} 		
\label{cp} 	
\end{figure}

Flow over an airfoil, NACA 0012 at different angles of attack $\alpha$ is considered. The computational domain consists of the contour of the airfoil embedded on a rectangular plane. The domain lengths upstream and downstream of the airfoil leading edge are 5C and 15C, respectively, and the domain width on either side of the leading edge is 5C. Here, C stands for the chord length.  The flow is set to be uniform at the inlet (the left boundary) and outflow boundary condition is applied at the outlet (the right boundary). The free stream boundary conditions are employed at the top and bottom boundaries of the domain, and a no-slip boundary condition is employed at the airfoil wall boundary.  	%%%%%%%%%%%%%%%%%
	Figure \ref{cp} shows the mean pressure distribution over the upper and lower surfaces of the airfoil at $Re = 1000$. The present result shows good agreement to those obtained by \cite{kurtulus2015unsteady}. 
	%%%%%%%%%%%%%  2  %%%%%%%%% 	
 	%%%%%%%%%%%%%%%%% 

\begin{figure} 		
\centering 		
\includegraphics[width=0.6\columnwidth]{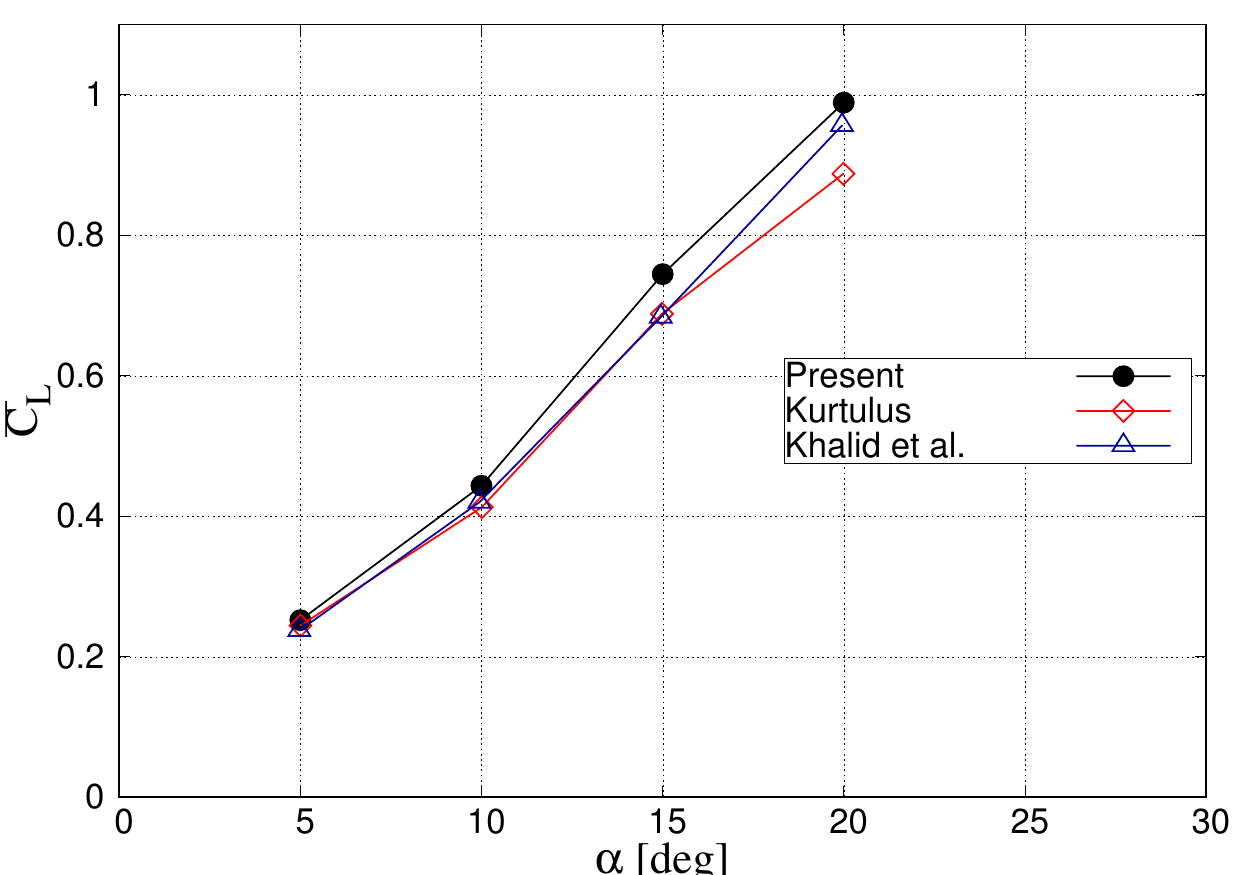} 		
\caption{Mean lift coefficient $\bar{C_L}$ as a function of angle of attack $\alpha$, for $Re = 1000$.} 		
\label{clmvsalpha} 	
\end{figure}

\begin{figure} 		
\centering 		
\includegraphics[width=0.6\columnwidth]{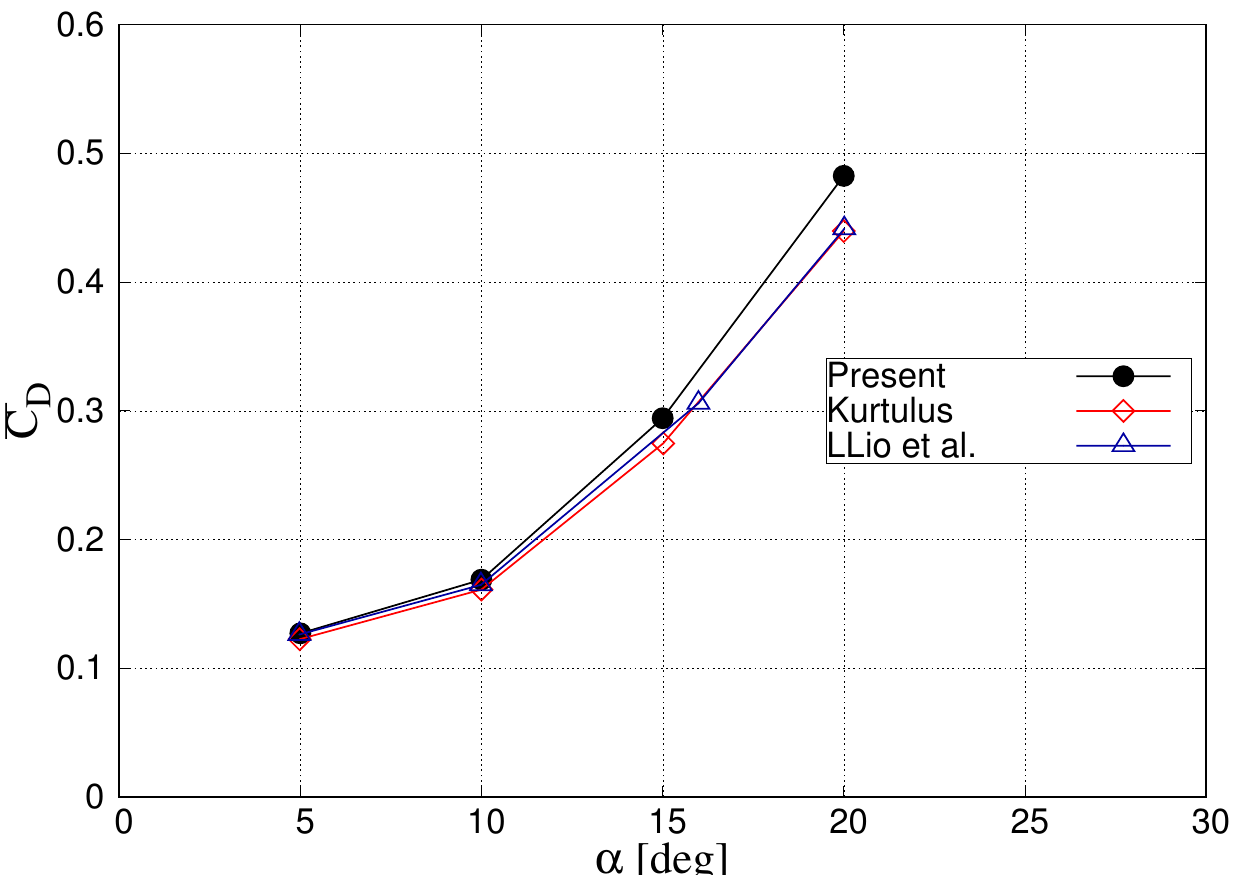} 		
\caption{Mean drag coefficient $\bar{C_D}$ as a function angle of attack $\alpha$, for $Re = 1000$.} 		
\label{cdmvsalpha} 	
\end{figure}

	The time-averaged lift coefficient $\bar{C_L}$ and drag coefficient $\bar{C_D}$ are shown in figures \ref{clmvsalpha} and \ref{cdmvsalpha}, respectively, for different angles of attack. The results show good agreement between the present and literature results at low angles of attack with reasonable difference of aerodynamic coefficients at $15^o$ and $20^o$ angle of attack. 	  	  	
%%%%%%%%%%%%%  3  %%%%%%%%% 	
 	%%%%%%%%%%%%%%%%% 
	%%%%%%%%%%%%%  4  %%%%%%%%% 	
 	%%%%%%%%%%%%%%%%% 

\subsection{Taylor-Green Vortices\label{subsec:Taylor-Green-Vortices}}

\begin{figure}
\centering{}\includegraphics[scale=0.4]{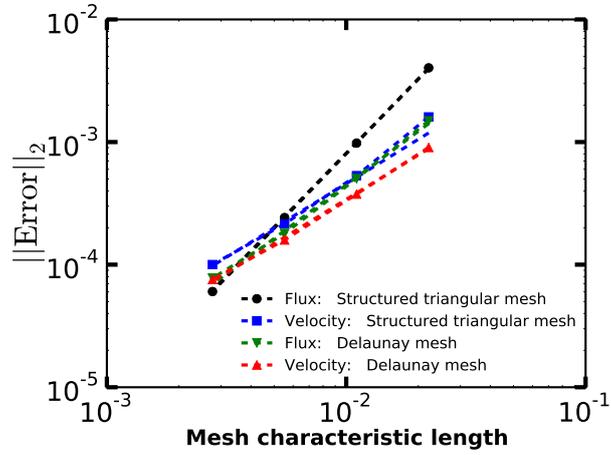}\caption{The convergence of the $L^{2}$-norm of the velocity 1-form (flux)
and the interpolated velocity vector errors with mesh size for the
Taylor-Green vortices case at time $t=10$. The dashed lines represent
the slopes of the orders: (i) black: 2.0, (ii) blue: 1.2, (iii) green:
1.45, (i) red: 1.2. \label{fig:The-convergence-of}}
\end{figure}

\begin{figure}
\subfloat[\label{fig:(a)}]{\centering{}\includegraphics[scale=0.17]{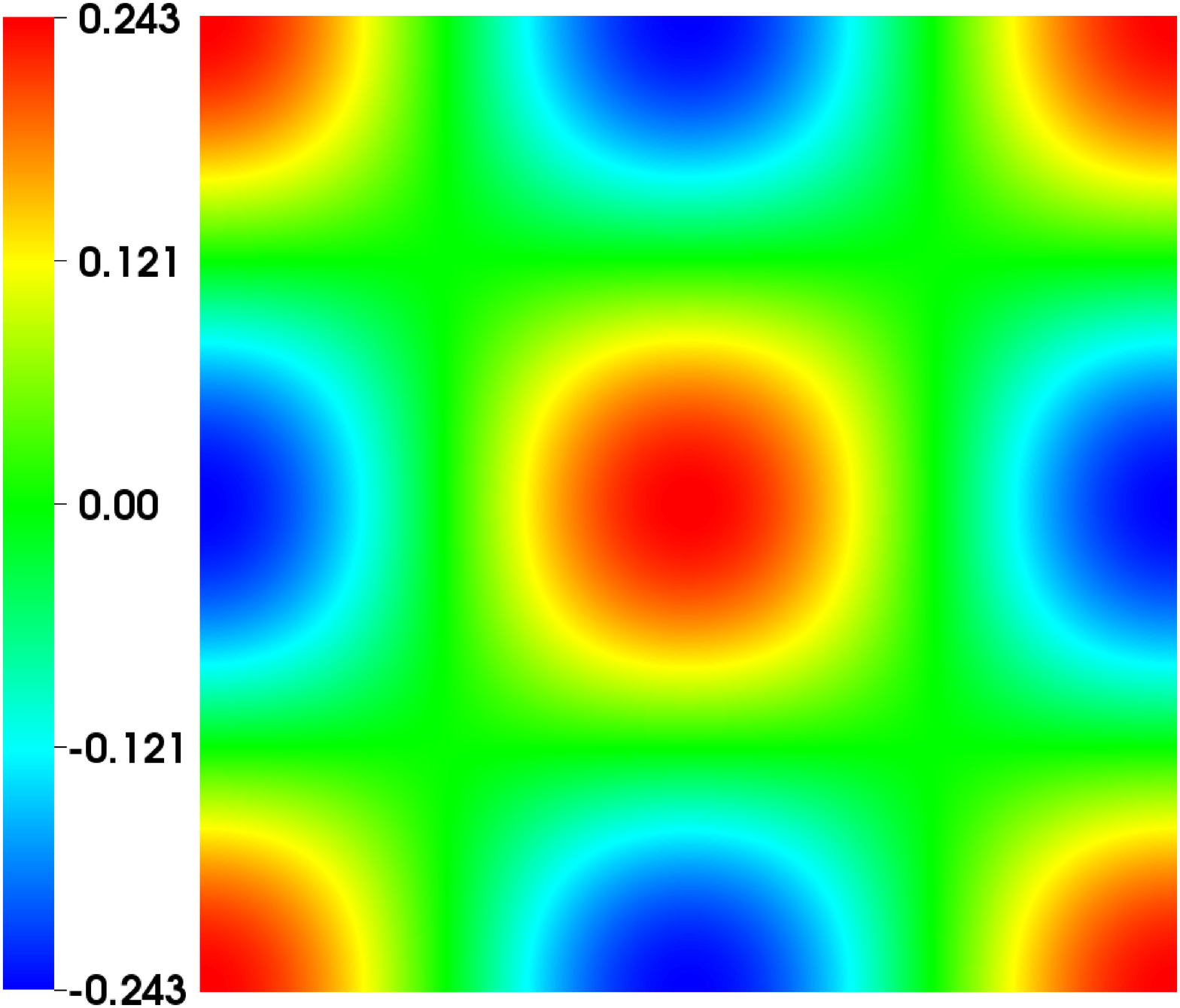}}\subfloat[\label{fig:(b)}]{\centering{}\includegraphics[scale=0.32]{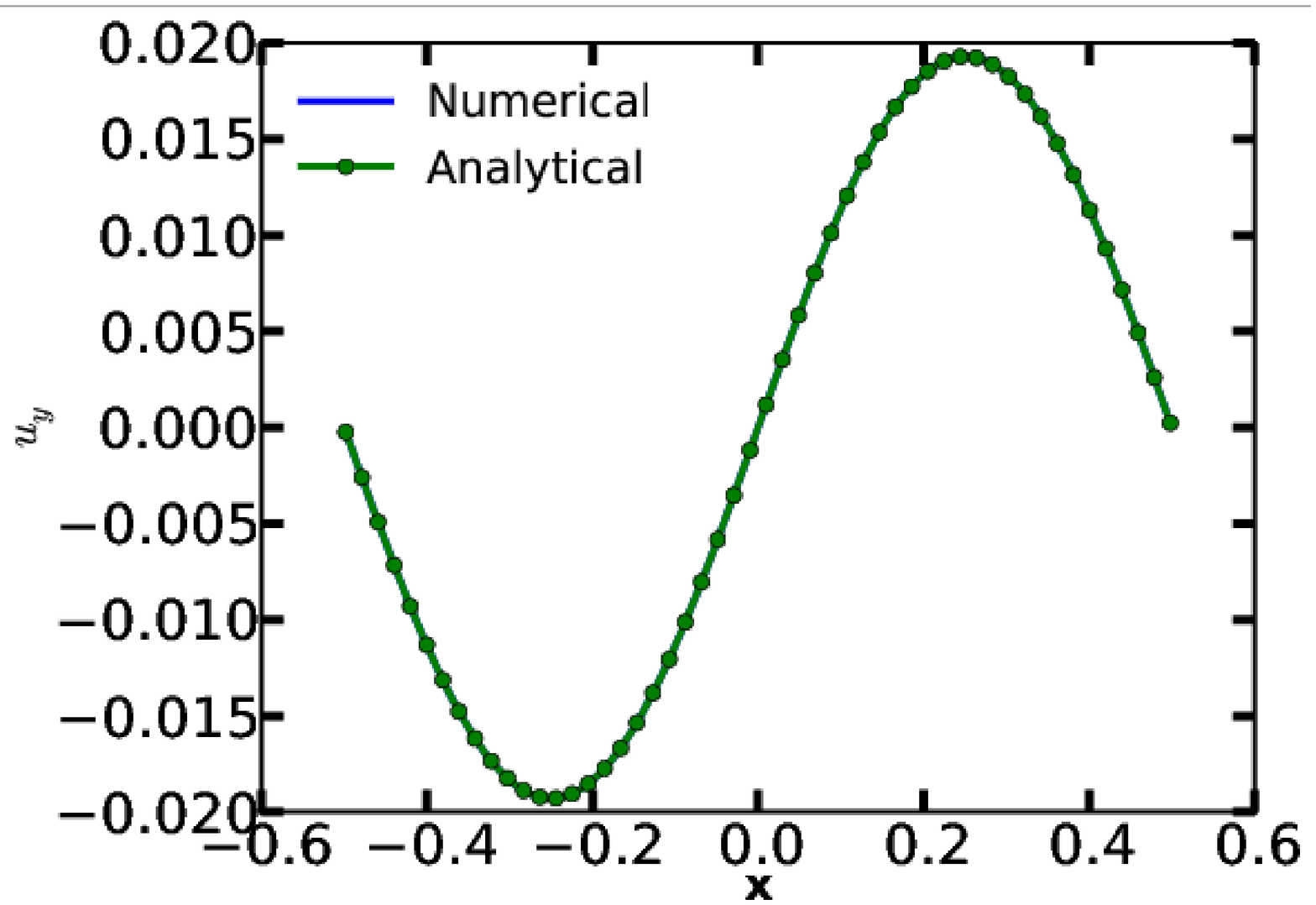}}\caption{(a) The vorticity distribution for Taylor-Green vortices at time $t=10$.
(b) The $y$-velocity distribution along the horizontal centerline
at time $t=10$.}
\end{figure}

The Taylor-Green vortices are simulated on a flat square domain of
dimension {[}-0.5,0.5{]} both in $x$ and $y$ coordinate directions.
The non-dimensional analytical solution for the decay of Taylor-Green vortices with
time in 2D reads \cite{connors2010convergence}

\begin{equation}
u_{x}=-\cos\left(2\pi x\right)\sin\left(2\pi y\right)e^{-8\pi^{2} t/Re},
\end{equation}

\begin{equation}
u_{y}=\sin\left(2\pi x\right)\cos\left(2\pi y\right)e^{-8\pi^{2} t/Re},
\end{equation}
where $u_{x}$ and $u_{y}$ are the $x$ and $y$ components of non-dimensional velocity
respectively, $x$ and $y$ are the non-dimensional coordinates, $Re$ is the Reynolds number, and $t$ is the non-dimensional time. The Reynolds number is defined as the ratio of viscous diffusion time to the circulation (turnover) time, i.e., $Re=\frac{L^{2}/\nu}{L/V}$. Here, $L$ is the characteristic length scale, $V$ is the characteristic velocity scale, and $\nu$ is the kinematic viscosity. The circulation time $\left(L/V\right)$ is used as the characteristic time scale.
At time $t=0$, the mass flux 1-form for a primal edge is determined from the integration of the normal component of velocity along the edge.
The Reynolds number  $Re=200$ 
is used for the simulations. Structured triangular meshes and Delaunay
meshes of different resolutions are used for the simulations and an
error convergence study is performed. Figure \ref{fig:The-convergence-of}
shows the convergence with the mesh element size of the $L^{2}$-norm
of the velocity 1-form (flux) error $\left\Vert  u ^{analytical}- u ^{numerical}\right\Vert =\left[\sum_{\sigma^{1}}\left(\frac{ u ^{analytical}- u ^{numerical}}{\left|\sigma^{1}\right|}\right)^{2}\left|\sigma^{1}\right|\left|\star\sigma^{1}\right|\right]^{1/2}$
\cite{mohamed2016discrete} at simulation time $t=10$. Here the superscripts
analytical and numerical stand for the analytical and numerical solutions,
respectively. We observe that the flux error converges with a second
order rate for the structured-triangular mesh, and with a rate somewhat
better than first order (i.e., $1.45$) otherwise, which agrees largely
with that reported in \cite{mohamed2016discrete}. Figure \ref{fig:The-convergence-of}
also shows the convergence of the $L^{2}$-norm of the velocity magnitude
error. The velocity vector field was interpolated from the flux distribution
using Whitney maps \cite{mohamed2016discrete}. The velocity error
converges with a rate slightly better than first order (i.e., $1.2$)
for both the structured triangular mesh as well as the Delaunay mesh.
For the structured triangular meshes, although the flux error converges
with a second order rate, the velocity error converges at a rate slightly
better than first order, due to the first order interpolation of velocity
field from the flux distribution. We note that the stream function
formulation \cite{mohamed2016discrete} also exhibits similar second
order flux error convergence rate for the structured-triangular meshes
and first order flux error convergence rate for the otherwise unstructured
meshes, and first order velocity error convergence rate for both
the structured-triangular meshes as well as the unstructured meshes.
For the post processing, the vorticity distribution (here, there is
only one component of vorticity normal to the domain plane) was determined
from the flux distribution as $\omega=\ast_{0}^{-1}\left[-\textrm{d}_{0}^{T}\right]\ast_{1} u ^{\ast}$.
Figure \ref{fig:(a)} shows the vorticity distribution at simulation
time $t=10$. Figure \ref{fig:(b)} shows, at simulation time $t=10$,
the distribution of the $y$-velocity along the horizontal domain
center line along with analytical solution, and the two agree well.

\subsection{Double Periodic Shear Layer\label{subsec:Double-Periodic-Shear}}

\begin{figure}
\centering{}\subfloat[\label{fig:t=00003D0}]{\centering{}\includegraphics[scale=0.63]{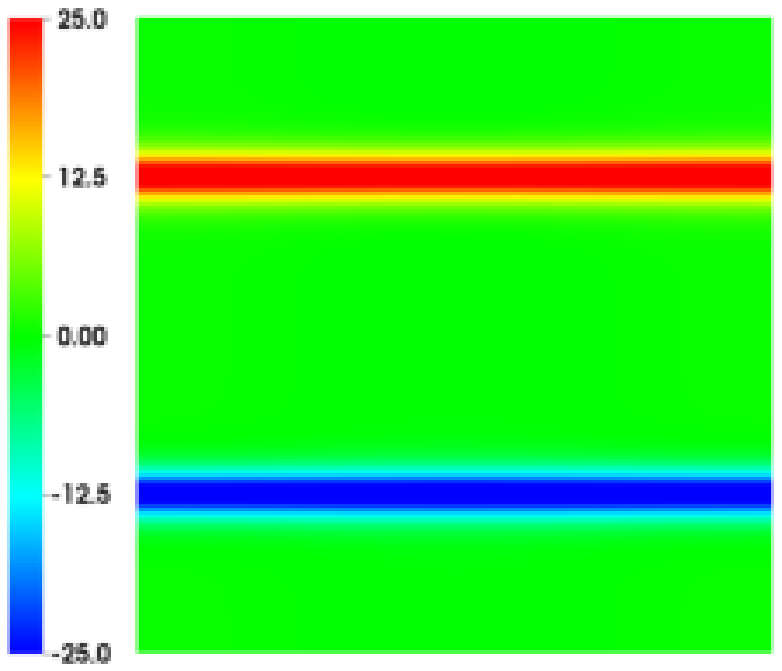}}\subfloat[\label{fig:t=00003D0.8}]{\centering{}\includegraphics[scale=0.63]{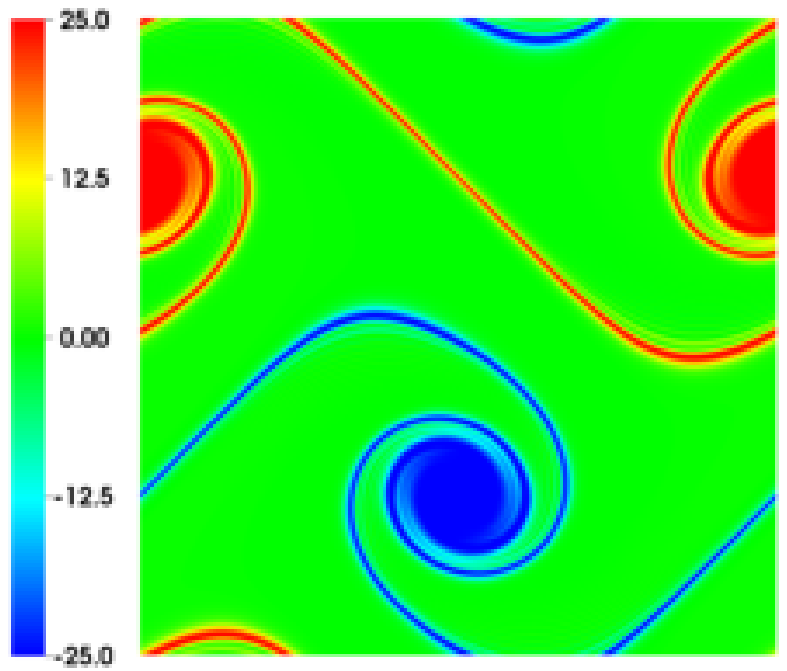}} \\ \subfloat[\label{fig:t=00003D1.2}]{\centering{}\includegraphics[scale=0.63]{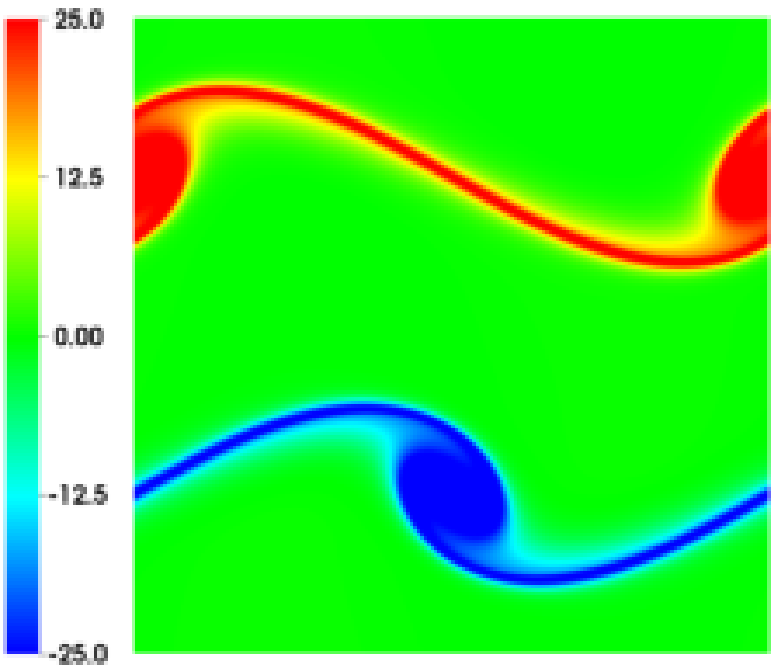}}  \subfloat[\label{fig:Kinetic-energy-relative}]{\centering{}\includegraphics[scale=0.3]{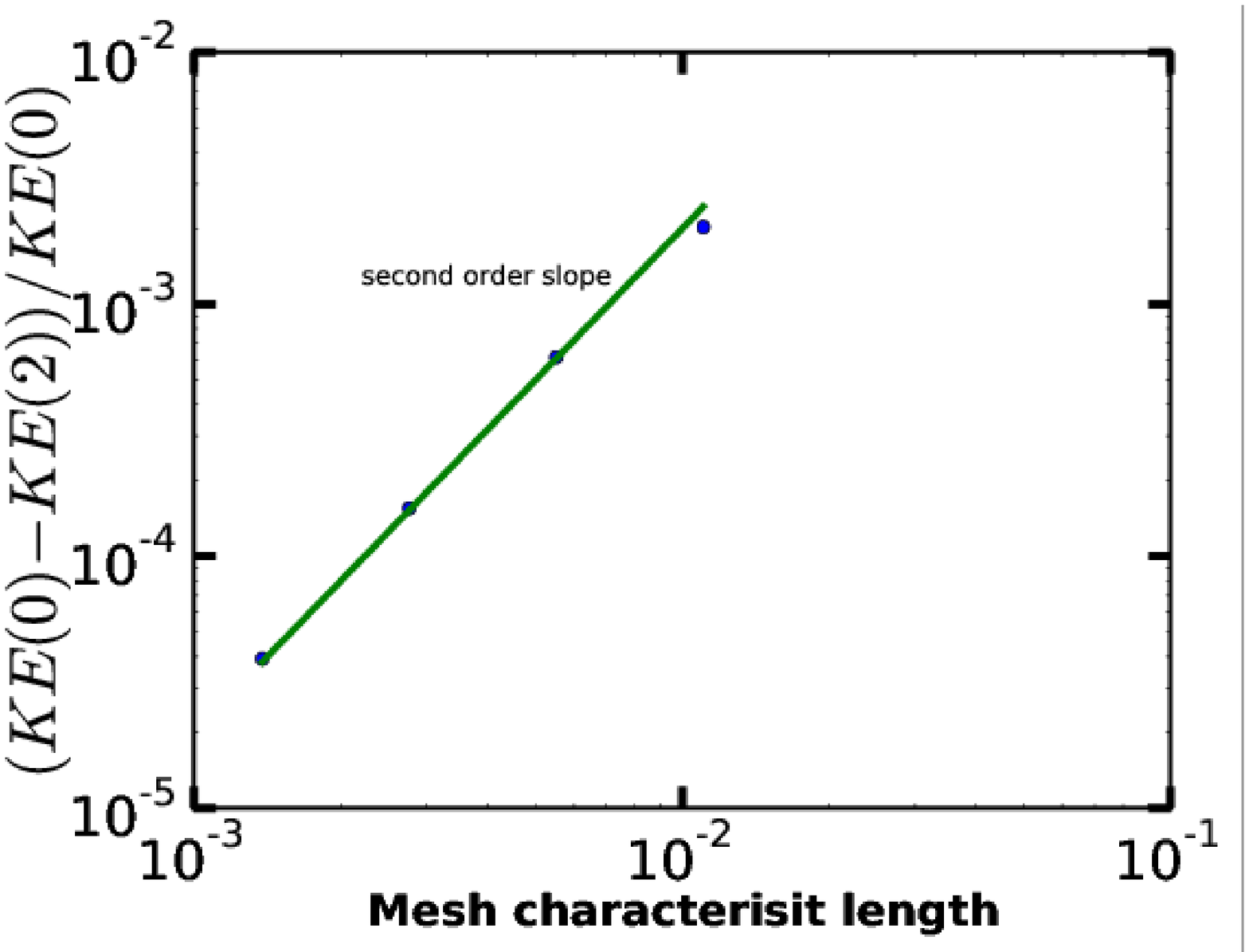}}\caption{Double periodic shear layer case. Vorticity magnitude distribution at (a) $t=0$, (b) $t=0.8$, (c) $t=1.2$. (d) kinetic energy relative error at $t=2$. \label{fig:Double-periodic-shear}}
\end{figure}

An inviscid flow and a square domain of unit edge length is assumed.
Figure \ref{fig:t=00003D0} shows the initial flow, which is a shear
layer of finite thickness with a small magnitude of vertical velocity
perturbation. The expression of the initial velocity field is \cite{bell1989second}

\begin{equation}
u_{x}=\begin{cases}
\begin{array}{c}
\tanh\left(\left(y-0.25\right)/\rho\right)\textrm{ for }y\leq0.5\\
\tanh\left(\left(0.75-y\right)/\rho\right)\textrm{ for }y>0.5
\end{array}\end{cases}
\end{equation}

\begin{equation}
u_{y}=\delta\sin\left(2\pi x\right)
\end{equation}
with $\rho=1/30$ and $\delta=0.05$. Similar to the case of Taylor-Green
vortices (section \ref{subsec:Taylor-Green-Vortices}), the initial
mass flux 1-forms are approximated from the integration of the normal component of velocity along the edges. Periodic
boundary conditions are imposed at all domain boundaries. Four mesh
resolutions consisting of structured-triangular meshes with number
of elements equal to $8192$, $32768$, $131072$ and $524288$ are
employed. A time step of $\Delta t=0.001$ is used. Figure \ref{fig:Double-periodic-shear}
shows the evolution of the vorticity field with time, using the finest
mesh. At time $t=0.8$, two well resolved vortices appear. Similar
to that reported in Mohamed et al. \cite{mohamed2016discrete}, the
shear layer connecting the coherent vortices becomes thinner with
time and within a finite time interval reach the resolution of the
mesh after which dispersion error manifests as mesh level oscillations.
The vorticity distribution plots in figure \ref{fig:Double-periodic-shear}
exhibit similarities with that reported previously by Mohamed et al.
\cite{mohamed2016discrete}, and Bell et al. \cite{bell1989second}.

Figure \ref{fig:Kinetic-energy-relative} the convergence of the kinetic
energy relative error $\left(\frac{\textrm{KE}\left(0\right)-\textrm{KE}\left(t\right)}{\textrm{KE}\left(0\right)}\right)$
with the mesh size at time $t=2.0$. Here, the total kinetic energy
$\textrm{\ensuremath{\left(\textrm{KE}\right)}}$ is calculated as
$\frac{1}{2}\int_{\Omega}\mathbf{v}\cdot\mathbf{v}d\Omega$, where
the integration is performed over the entire simulation domain, where
the velocity vector $\left(\mathbf{v}\right)$ is calculated in each
triangular element via Whitney map interpolation, as discussed before.
As expected, the kinetic energy relative error converges in a second
order fashion with the mesh size. It is worth noting that the stream
function formulation \cite{mohamed2016discrete} also exhibits similar
second order convergence rate of the kinetic energy relative error.
Kinetic energy relative error is $0.2\%$ for the coarsest mesh (equivalent
to $64\times64$ Cartesian mesh) and it is only $0.0039\%$ for the
finest mesh (equivalent to $512\times512$ Cartesian mesh). It is
$0.06\%$ for the mesh with $32768$ triangular elements (equivalent
to a $128\times128$ Cartesian mesh), almost one order of magnitude
lower than a second order collocated mesh scheme \cite{bell1989second}
using almost the same mesh size.

\subsection{Motion of Harmonic Waves on a Rotating Sphere\label{subsec:Motion-of-Harmonic}}

\begin{figure}
\subfloat[\label{fig:t=00003D0-1}]{
\begin{centering}
\includegraphics[scale=0.9]{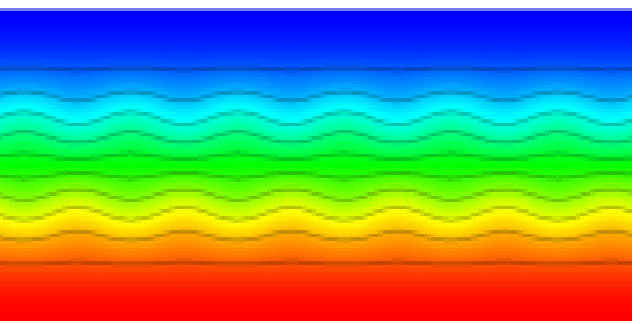}
\par\end{centering}
}\subfloat[\label{fig:t=00003D36.39-days}]{
\centering{}\includegraphics[scale=0.9]{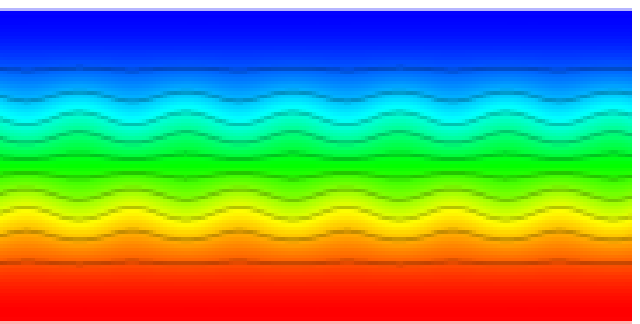}}\caption{Stream function distribution for the motion of harmonic waves on a
rotating sphere case. (a) $t=0$, (b) $t=36.39 $ days. \label{fig:Stream-function-distribution}}
\end{figure}

There exists a stationary solution to the motion of harmonic waves
in the Earth's atmosphere. These waves are the so-called Rossby waves.  
The stationary solution is \cite{neamtan1946motion}

\begin{equation}
\psi\left(\phi,\theta\right)=A\sin m\phi P_{l}^{m}\left(\cos\theta\right)-BR^{2}\cos\theta,
\end{equation}
where, $\psi$ is the stream function, $B=\frac{2\Omega}{l\left(l+1\right)-2}$,
$\phi$ is longitude, $\theta$ is the colatitude
angle, $A$ is an arbitrary constant, $P_{l}^{m}$
is associated Legendre polynomial of degree $l$ and order $m$, $R$
is earth's radius, and $\Omega$ is the rate of earth's rotation. An inviscid flow and $f=2\Omega\cos\theta$ are considered. 
The stream function distribution is shown in figure \ref{fig:Stream-function-distribution}
for $A=200$, $l=7$, $m=6$, $R=6.371\times10^{6}$, and $\Omega=7.2921\times10^{-5}$.
Figure \ref{fig:t=00003D0-1} shows the stationary solution at time
$t=0$, and \ref{fig:t=00003D36.39-days} shows the one at time $t=3.144\times10^{6}$
s or $36.39$ days. There is no distortion of the stream function
distribution and the stationary solution is preserved. However, there
is a westward phase displacement of $28.6^{\circ}$ longitude after
$36$ days. ($5.7^{\circ}$ longitude after 7 days). Similar westward
phase displacement of $7^{\circ}$ longitude after $8$ days has been
reported in \cite{sadourny1968integration}.

\begin{figure}
\subfloat[]{
\begin{centering}
\includegraphics[scale=0.3]{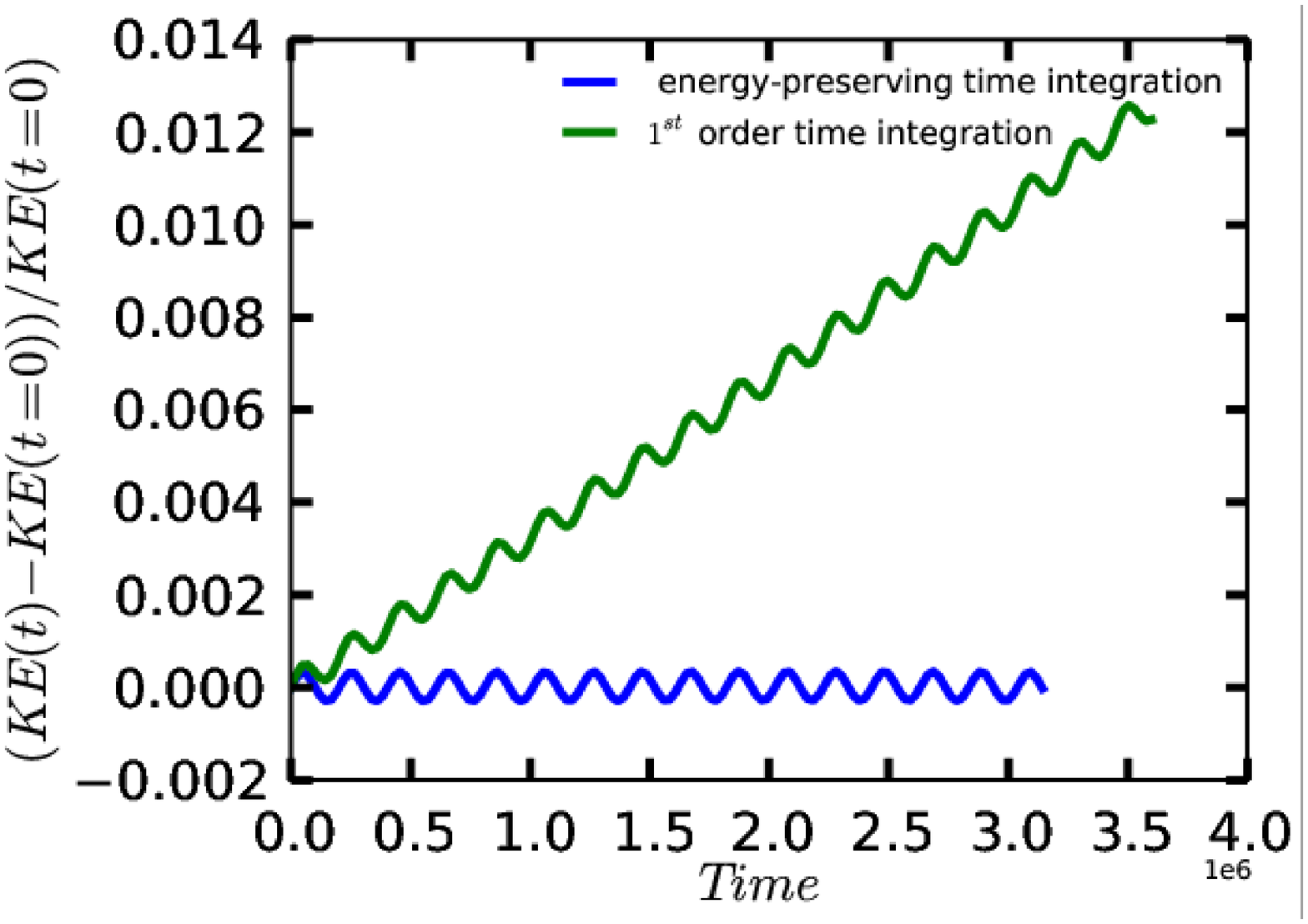}
\par\end{centering}
}\subfloat[]{
\begin{centering}
\includegraphics[scale=0.3]{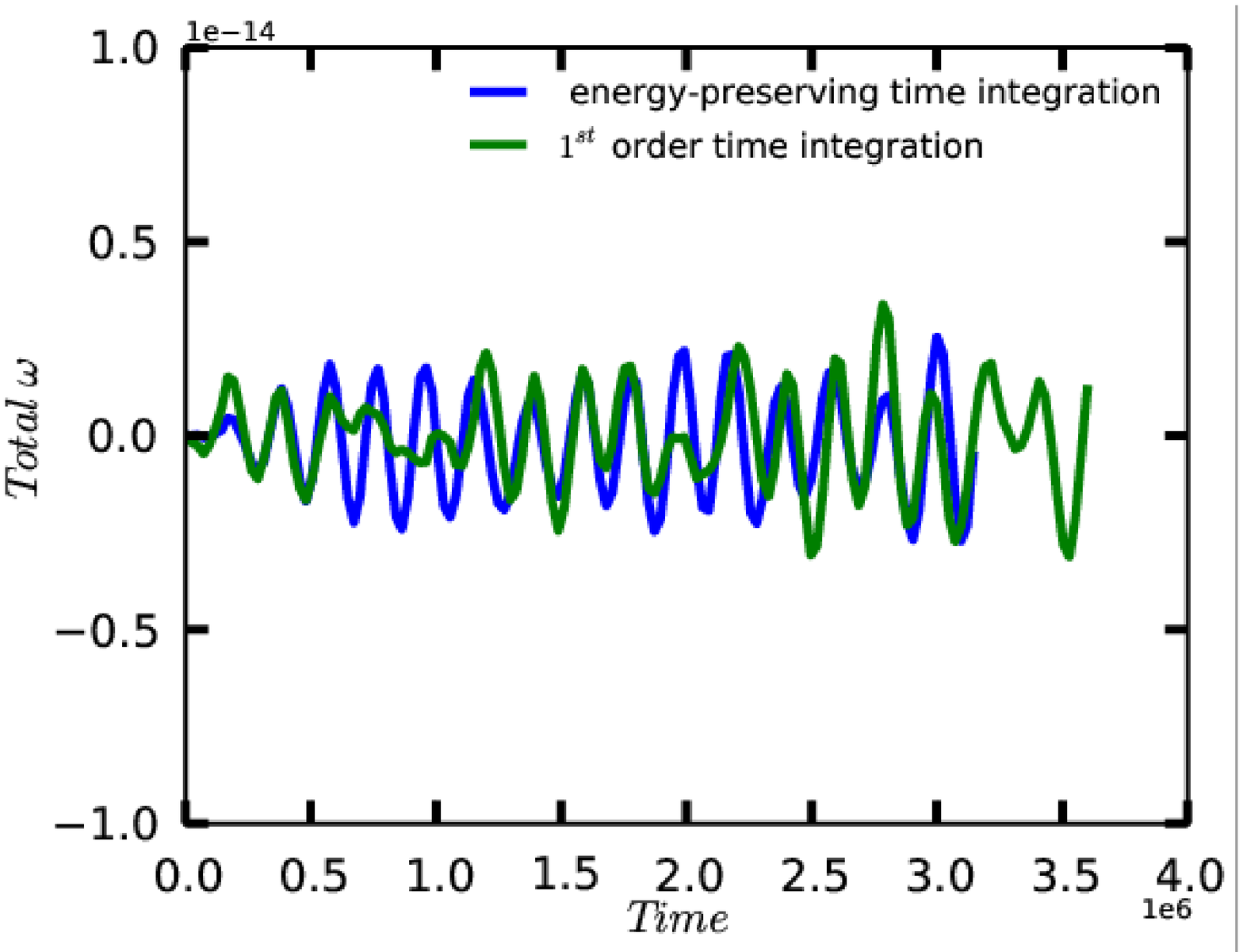}
\par\end{centering} 
} \\ \subfloat[]{
\begin{centering}
\includegraphics[scale=0.3]{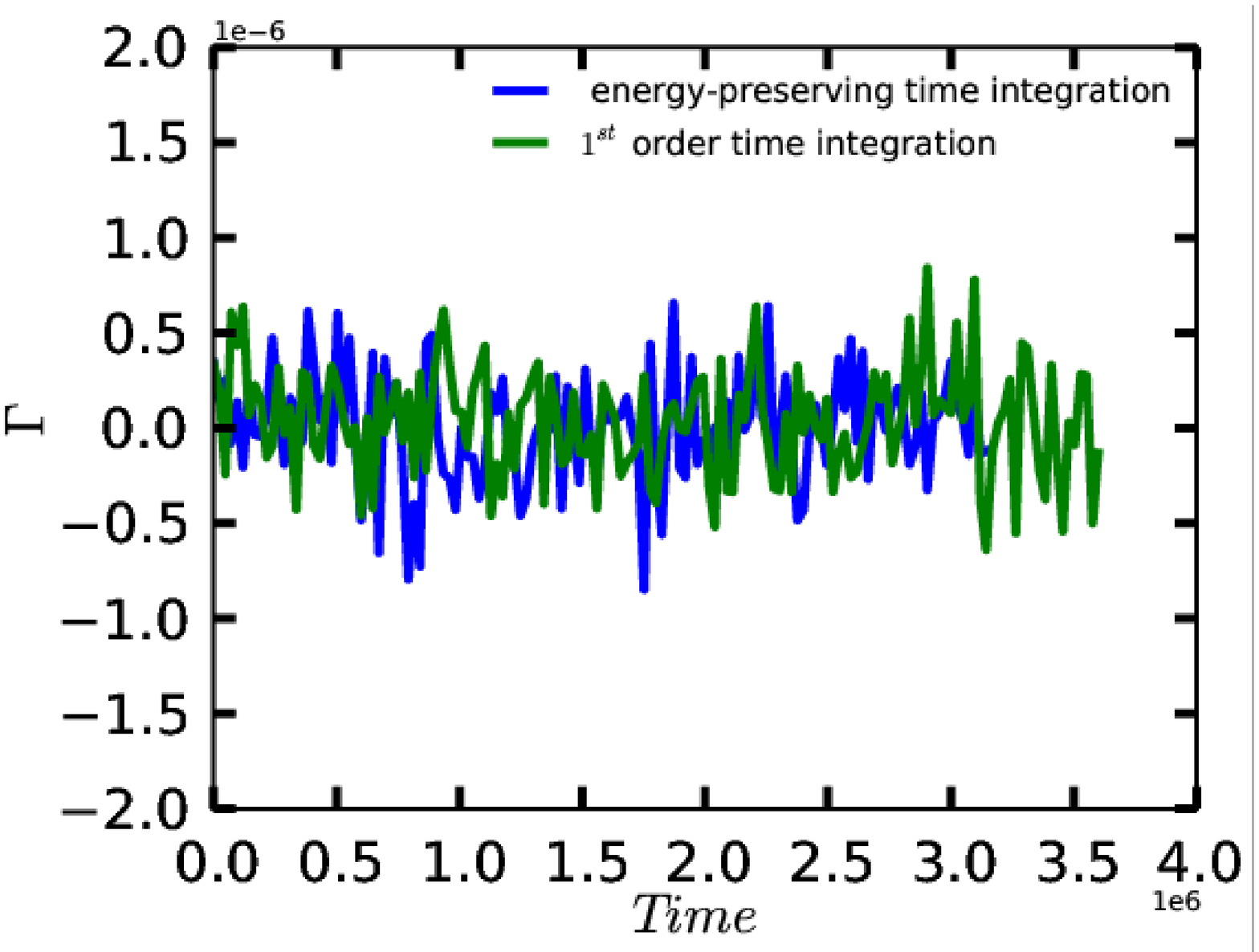}
\par\end{centering}
}\subfloat[]{
\begin{centering}
\includegraphics[scale=0.3]{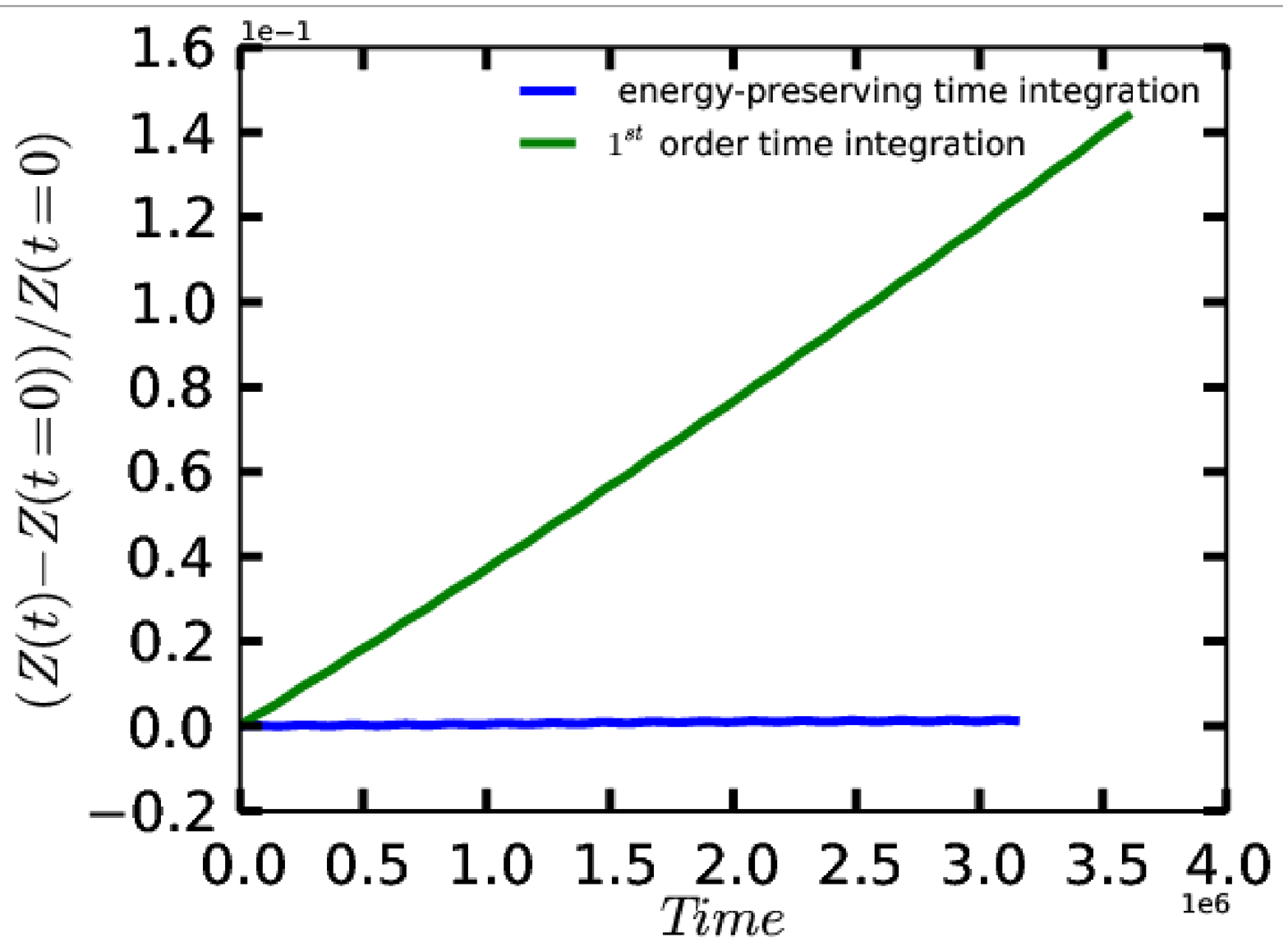}
\par\end{centering}
}\caption{Inviscid invariants as a function of time for the motion of harmonic
waves on a rotating sphere case. (a) Total relative kinetic energy, (b) Total vorticity, (c) Total circulation, (d) Total relative enstrophy. \label{fig:Inviscid-invariants-as}}
\end{figure}

Figure \ref{fig:Inviscid-invariants-as} shows various inviscid invariants
as a function of time, and demonstrates conservation of these invariants
for an extended period of time. It is observed that energy-preserving
time integration has a superior conservation of total kinetic energy
and the total enstrophy as compared to the Euler first order time
integration. The total kinetic energy is already defined in section
\ref{subsec:Double-Periodic-Shear}. The total enstrophy is calculated
as $\frac{1}{2}\int_{\Omega}\omega^{2}d\Omega$, where the integration
is performed over the entire simulation domain, and $\omega$ is
the vorticity. 

\section{Conclusions}

We presented a primitive variable DEC formulation of Navier-Stokes
equations on surface simplicial meshes. Here, the mass flux 1-forms
at all mesh primal edges, and the dynamic pressure 0-forms at the
mesh dual nodes (triangle circumcenters) are the degrees of freedom.
The discrete operators emerging from the DEC discretization are discussed.
Two time integration schemes, i.e., Euler first order, and energy-preserving,
are implemented. The Coriolis force is incorporated to facilitate
the use of the method to study flows on rotating surfaces. The method
exhibits second order flux error convergence rate with the mesh size for the structured triangular meshes, and a rate somewhat
better than first order (i.e., 1.45) for unstructured meshes. Moreover,
the method exhibits slightly better than first order (i.e., 1.2) error
convergence rate with the mesh size for the interpolated velocity vector for both the structured triangular as well as unstructured meshes. The
method exhibits second order kinetic energy relative error convergence
rate with mesh size for inviscid flows. The method preserves a stationary
state for the flow on a rotating sphere over an extended period of
time. The energy-preserving time integrator exhibits superior conservation
of inviscid invariants such as total kinetic energy and total enstrophy
for an extended period of time.

\section*{References}

\bibliography{paper}

\appendix
\section*{Appendix}
\subsection*{Primer on Discrete Exterior Calculus (DEC)\label{sec:Primer-on-Discrete}}
The calculus of {\em differential forms} is dubbed exterior calculus. While vector calculus deals with scalars, vectors and tensors, exterior calculus deals with differential forms ($k$-forms). Perot et al. \cite{perot2014differential} describes differential forms, with a view point of their applications to science and engineering, and is largely referred to in the ensuing discussion. A differential form can be regarded as ``the things which occur under integral signs"  \cite{flanders1963differential}. For example, for a line integral of a vector $ u $, $\int u \cdot\mathbf{dl}$, $ u \cdot\mathbf{dl}$ is a 1-form $u^{1}$. Similarly, a 2-form $a^{2}=\mathbf{a}\cdot\mathbf{dA}$ appears in surface integrals, a 3-form $b^{3}=bdV$ appears in volume integrals, and since a 0-integral is trivial, a 0-form $p^{0}$ is a scalar function. Here, the superscript on the form is the dimension of the form. On the other hand, the discrete forms are the entire integral quantities. For example, integration of a vector $ u $ over a mesh edge $\int u \cdot\mathbf{dl}$ is a discrete 1-form $u^{1}$. Similarly, surface integral $\int\mathbf{a}\cdot\mathbf{dA}$ over a mesh face represents a discrete 2-form $a^{2}$, and a discrete 3-form is the volume integral $b^{3}=\int bdV$ on a cell volume. Since the integration over a point is trivial and it just returns the value of the integrand at that point, a discrete 0-form is just a scalar field value at a mesh vertex/node. The perception of discrete 0-form and that of the discrete 1-form remain essentially the same for two-dimensional (2D) and three-dimensional (3D) spaces. However, since the mesh cells are represented by areas in 2D, the discrete 2-form is a cell average integral of a scalar field in 2D rather than integral of a vector field on a mesh face as in 3D. The confining space dimension is the highest dimension of a form. Therefore, there does not exist a 3-form in 2D, and the 2-form essentially behaves like the 3-form does in 3D.
Other key concepts are the {\em exterior derivative} $\left(\textrm{d}\right)$, the {\em Hodge star operator} $\left(\ast\right)$, and  the {\em exterior product or wedge product} $\left(\wedge\right)$. The exterior derivative implies multidimensional differentiation, and is a generalization of the usual curl, divergence and gradient operations of calculus. It operates on a form and results in to the next higher dimensional form. The gradient of a scalar in 3D, i.e., $\nabla q=\mathbf{p}$ in vector calculus is equivalent to $\textrm{d}q^{0}=p^{1}$ in exterior calculus. Here, the 1-form $p^{1}$ best represents the resulting vector $\mathbf{p}$. The curl of a vector, $\nabla\times u =\mathbf{a}$ is equivalent to $\textrm{d}u^{1}=a^{2}$. The resulting vector $\mathbf{a}$ (in vector calculus) is represented as a 2-form (in exterior calculus).  The divergence of a vector, $\nabla\cdot u =b$ is equivalent to $\textrm{d}u^{2}=b^{3}$, resulting in to a 3-form which is a scalar. In 2D, gradient $\nabla,$ and 2D curl $\nabla\times_{2D}$ are the only primary differentiation operations, and the divergence does not add any useful content. The gradient of a scalar looks the same as in 3D, i.e., it is equivalent to the exterior derivative operating on a 0-form,  and implies the difference between two node values / discrete 0-forms. The curl of a vector lying in a plane , $\nabla\times_{2D} u =\frac{\partial u_{y}}{\partial x_{x}}-\frac{\partial u_{x}}{\partial x_{y}}$, is equivalent to a 2-form resulting from the application of an exterior derivative to a 1-form (on an edge). The discrete exterior derivative $\textrm{d}_0$ operates on  primal 0-forms and results in to  primal 1-forms, and the discrete exterior derivative $\textrm{d}_1$ operates on primal 1-forms and results in to  primal 2-forms. Similarly, (in 2D) the discrete exterior derivative $\left[-\textrm{d}_0^{T}\right]$ (the negative sign is due to the adopted mesh orientation convention) operates on  dual 0-forms and results in to  dual 1-forms and the discrete exterior derivative $\textrm{d}_1^{T}$ operates on dual 1-forms and results in to  dual 2-forms. The Hodge star operator has a metric term incorporated in it. The Hodge star operating on a form in one dimension (e.g., $k$) results in to a form in the reflected dimension ($N-k$), i.e., $\ast a^{k}=b^{N-k}$, where $N$ stands for the space dimension. In 3D, Hodge star on a 0-form (scalar value at a point / node) results in to a 3-form (cell average of scalar), and vice-versa. Applying Hodge star to a 1-form (vector component along a line) results in a 2-form (vector component normal to a face) and vice-versa. In practice, this implies that a discrete Hodge star operator transfers the forms on a mesh to a dual mesh (as well as changes their dimension). One need to identify which dual mesh is being used for the definition of the discrete Hodge star matrices, since every well formed mesh has an infinite number of well formed dual meshes. The median dual mesh (barycentric dual mesh), and the voronoi dual mesh (circumcentric dual mesh) are the examples of dual meshes for unstructured (triangular or tetrahedral) primary meshes.  The discrete Hodge star $\left(\ast_{0}\right)$ operating on cell node (vertex) values (0-forms) of a tetrahedral mesh results in cell average values (dual 3-forms) for the surrounding Voronoi cells. Similarly, the inverse operator $\left(\ast_{1}^{-1}\right)$ operates on 1-forms lying on a dual mesh edge (i.e., on the line between circumcenters of two adjacent primal cells), and results in to 2-forms on the primal mesh (normal flux values on the cell faces). In 2D, a discrete $\ast_{0}$ operates on mesh vertex values (primal 0-forms), and results in to Voronoi cell values (dual 2-forms). The operation $\ast_{0}^{-1}$ transfers the cell average values (dual 2-forms) to mesh vertex values (primal 0-forms). The discrete $\ast_{1}$ operator transfers the primal 1-forms (primal edge values) to dual 1-forms (dual edge values). On the other hand, The discrete $\ast_{1}^{-1}$ operator transfers the dual 1-forms (dual edge values) to primal 1-forms (primal edge values). The wedge product of two forms implies multiplication of the forms. The dimension of the resultant form is the sum of the dimensions of the two forms in the wedge product, i.e., $a^{j}\wedge b^{k}=c^{j+k}$ for $j+k\leq N$, with a property $a^{j}\wedge b^{k}=\left(-1\right)^{jk}b^{k}\wedge a^{j}$. The wedge product is also associative, i.e., $\left(a^{j}\wedge b^{k}\right)\wedge c^{l}=a^{j}\wedge\left(b^{k}\wedge c^{l}\right)$ for $j+k+l\leq N$. In 3D, the wedge product implies a regular multiply, a dot-product, or a cross-product depending on the dimension of the two forms involved in the wedge product. The wedge product implies a regular multiplication, if either operand is a scalar 0-form. When one of the operands is a 1-form and the other is a 2-form the wedge product implies a dot product of them. The wedge product of two 1-forms implies cross product on the vector proxies of the forms, resulting in a 2-form. In 2D, the wedge product of two forms implies either a regular multiplication or a 2D cross product of the forms. Similar to 3D, if either operand is a 0-form, a regular multiply is implied. For the two forms $a^{1}$ and $b^{1}$ associated with the vectors $\mathbf{a}$ and $\mathbf{b}$, respectively, the wedge product $a^{1}\wedge b^{1}$ is equivalent to $\mathbf{a}\times_{2D}\mathbf{b}=a_{x}b_{y}-a_{y}b_{x}$. 
Using integral quantities / discrete forms as the primary unknowns for a numerical method transforms any continuous PDE system in to a discrete algebraic system such that all errors/approximations occur at the algebraic level only and not in the approximation of the differential operators of the PDE \cite{perot2007discrete}. Thus, DEC can mimic physical and mathematical properties of the continuous PDE system. Moreover, DEC being coordinate independent, it is a more convenient numerical tool to investigate flows over arbitrary curved surfaces/manifolds. The DEC implementation requires a primal simplicial complex (triangular mesh in 2D or tetrahedral mesh in 3D), along with a dual mesh associated to it (usually circumcentric). In 2D, the primal simplicial complex consists of the set of nodes, edges, and triangles, and the dual mesh consists of dual polygons, edges and nodes associated with the primal mesh objects. DEC discretization of a physical problem expresses the physical fields discretely as scalars representing the integration of these $k$-forms on $k$-dimensional primal/dual mesh objects as already mentioned above. A few representative DEC references include \cite{flanders1963differential, abraham2012manifolds, hirani2003discrete, desbrun2003discrete, desbrun2005discrete, grinspun2006discrete, hirani2008numerical, desbrun2008discrete, crane2013digital, perot2014differential, mohamed2016discrete, de2016subdivision}. 

\end{document}